\documentclass[12pt]{article}
\usepackage{a4wide}
\usepackage{latexsym}
\usepackage{amsmath}
\usepackage{amsfonts}
\usepackage{amsthm}
\usepackage{amscd}
\usepackage{cite}

\usepackage{pslatex}
\usepackage{graphicx}
\usepackage[latin1,utf8]{inputenc}
\usepackage[T1]{fontenc}

\usepackage{ifthen}

\allowdisplaybreaks

\newcommand{\bq}{\begin{eqnarray}}
\newcommand{\eq}{\end{eqnarray}}
\newcommand{\eps}{\varepsilon}

\newcommand{\slashoperator}[2]{|_{#2} #1}

\newtheorem{Def}{Definition}[section]
\newtheorem{Theo}{Theorem}[section]

\newboolean{arxiveversion}
\setboolean{arxiveversion}{true}

\ifthenelse{\boolean{arxiveversion}}
{
} 
{
} 

\begin{document}

\thispagestyle{empty}

\begin{flushright}
  MITP/17-027
\end{flushright}

\vspace{1.5cm}

\begin{center}
  {\Large\bf Feynman integrals and iterated integrals of modular forms \\ 
  }
  \vspace{1cm}
  {\large Luise Adams and Stefan Weinzierl \\
  \vspace{1cm}
      {\small \em PRISMA Cluster of Excellence, Institut f{\"u}r Physik, }\\
      {\small \em Johannes Gutenberg-Universit{\"a}t Mainz,}\\
      {\small \em D - 55099 Mainz, Germany}\\
  } 
\end{center}

\vspace{2cm}

\begin{abstract}\noindent
  {
In this paper we show that certain Feynman integrals can be expressed 
as linear combinations of iterated integrals of modular forms
to all orders in the dimensional regularisation parameter $\varepsilon$ .
We discuss explicitly the equal mass sunrise integral and the kite integral.
For both cases we give the alphabet of letters occurring in the iterated integrals.
For the sunrise integral we present a compact formula, expressing this integral to all orders in $\varepsilon$
as iterated integrals of modular forms.
   }
\end{abstract}

\vspace*{\fill}

\newpage

\section{Introduction}
\label{sec:intro}

Precision predictions in high-energy particle physics rely on our ability to compute higher-order terms in perturbation theory
for scattering amplitudes.
In particular one faces the challenge to calculate quantum loop amplitudes, involving Feynman integrals.
Unfortunately we are far away from having a complete theory telling us to which functions Feynman integrals evaluate.
The best we can do is to analyse step-by-step classes of Feynman integrals of increasing sophistication.
Fortunately, several methods from mathematics -- and here from algebraic geometry in particular -- have entered in recent years
the field of precision calculations in particle 
physics \cite{Bloch:2005,Bloch:2008jk,Bogner:2007cr,Bogner:2007mn,Smirnov:2010hn,Schabinger:2011dz,Chen:2015lyz,Kosower:2011ty,CaronHuot:2012ab,Badger:2012dv,Lee:2013hzt,Sogaard:2014jla,Larsen:2015ped,Georgoudis:2016wff,Larsen:2017aqb,Bosma:2017ens,Frellesvig:2017aai,Ita:2015tya,Abreu:2017idw,vonManteuffel:2014ixa,Peraro:2016wsq}.

Multiple polylogarithms \cite{Goncharov_no_note,Goncharov:2001,Borwein}
play an important role for a wide class of Feynman integrals.
The richness of their mathematical structure derives from the fact that they have 
at the same time
a representation in the form of nested sums 
and
a representation in the form of iterated integrals.
Over the years many techniques and algorithms have been developed to evaluate Feynman integrals from this class
to multiple polylogarithms 
\cite{Vermaseren:1998uu,Remiddi:1999ew,Moch:2001zr,Weinzierl:2002hv,Weinzierl:2004bn,Moch:2005uc,Bierenbaum:2003ud,Brown:2008,Panzer:2014caa,Bogner:2015unknown,Ablinger:2015tua,Kotikov:1990kg,Kotikov:1991pm,Remiddi:1997ny,Gehrmann:1999as,Argeri:2007up,MullerStach:2012mp,Henn:2013pwa,Henn:2014qga,Tancredi:2015pta,Primo:2016ebd,Adams:2017tga,Gehrmann:2014bfa,Argeri:2014qva,Lee:2014ioa,Prausa:2017ltv,Gituliar:2017vzm,Meyer:2016slj}.

However, it is well-known that the class of functions of multiple polylogarithms is not large enough to express all 
Feynman integrals.
Prominent examples of Feynman integrals which cannot be expressed in terms of multiple polylogarithms are
the sunrise integral \cite{Broadhurst:1993mw,Berends:1993ee,Bauberger:1994nk,Bauberger:1994by,Bauberger:1994hx,Caffo:1998du,Laporta:2004rb,Kniehl:2005bc,Groote:2005ay,Groote:2012pa,Bailey:2008ib,MullerStach:2011ru,Adams:2013nia,Bloch:2013tra,Adams:2014vja,Adams:2015gva,Adams:2015ydq,Remiddi:2013joa,Bloch:2016izu}
or the kite integral\cite{Sabry:1962,Remiddi:2016gno,Adams:2016xah}.
Further examples from quantum field theory can be found in \cite{Bloch:2014qca,Primo:2017ipr,Bonciani:2016qxi,vonManteuffel:2017hms}, related examples from string theory are discussed in \cite{Broedel:2014vla,Broedel:2015hia,Broedel:2017jdo,D'Hoker:2015qmf,Hohenegger:2017kqy}.
Current research efforts are centred around suitable generalisations of multiple polylogarithms.
Common to all examples of Feynman integrals evaluating beyond the class of multiple polylogarithms and discussed so far 
is the fact that their geometry is related to an elliptic curve.
The extension of the class of functions required to express these integrals
consisted in a generalisation of the multiple polylogarithms towards the 
elliptic setting \cite{Beilinson:1994,Levin:1997,Levin:2007,Enriquez:2010,Brown:2011,Wildeshaus,Bloch:2013tra,Bloch:2014qca,Adams:2014vja,Adams:2015gva,Adams:2015ydq,Adams:2016xah,Passarino:2017EPJC}.
Multiple polylogarithms are associated to a punctured Riemann surface of genus zero \cite{Brown:2006}
and elliptic generalisations of multiple polylogarithms are associated to a punctured Riemann surface 
of genus one \cite{Brown:2011}.
In \cite{Adams:2015ydq,Adams:2016xah} we were able to express the equal mass sunrise integral and the kite
integral to all orders in the dimensional regularisation parameter in terms of elliptic generalisations
of (multiple) polylogarithms.
The concrete set of functions 
which we used ($\mathrm{ELi}$-functions, reviewed in appendix~\ref{sect:ELi})
were obtained from generalising the sum representation of the (multiple) polylogarithms towards the elliptic setting.

The motivation for this paper is two-fold: First we would like to investigate in more detail
the iterated integral aspects of our results.
Secondly we would like to explore the modular properties of the solutions.
The outcome of these investigations rewards our efforts:
We show that the equal mass sunrise integral and the kite integral can be expressed to all orders
as iterated integrals of modular forms.
For the sunrise integral the alphabet consists of three modular forms $f_2$, $f_3$ and $f_4$ 
of modular weight $2$, $3$ and $4$, respectively and
the trivial constant modular form $1$ (of modular weight $0$).
For the kite integral we need on top of those five additional modular forms, three modular forms of weight $2$
(which we label $g_{2,0}$, $g_{2,1}$ and $g_{2,9}$) and two modular forms of weight $3$ (labelled 
$g_{3,0}$ and $g_{3,1}$).
All of these modular forms are modular forms of a congruence subgroup.
The concrete congruence subgroup depends on the elliptic curve we start with and on the choice of periods for the elliptic curve.
We present two calculations, 
one where we start from an elliptic curve obtained 
from the second graph polynomial of the sunrise integral 
in the Feynman parameter representation.
In the Euclidean region the lattice generated by the periods of this elliptic curve is rectangular 
and we may choose the periods such that
one period is real, the other purely imaginary in the Euclidean region.
With this choice the congruence subgroup turns out to be $\Gamma_1(12)$.
It is possible to express all formulae in terms of modular forms from the smaller space
of the congruence subgroup $\Gamma_1(6)$.
One possibility to arrive at $\Gamma_1(6)$ is to start from an elliptic curve obtained from the maximal cut of the sunrise integral.
This is our second calculation. In this case we no longer have a rectangular lattice in the Euclidean region.
However, with a standard choice of periods we obtain directly the congruence subgroup $\Gamma_1(6)$.
Of course, both approaches give identical results and we discuss the relation between the two approaches.
The two elliptic curves are related by a quadratic transformation.
and the final formulae are related by a simple substitution $q_C=-q_2$.

Let us stress that for a different choice of periods one also obtains $\Gamma_1(6)$ when working with the elliptic curve
obtained from the second graph polynomial of the sunrise integral 
in the Feynman parameter representation.
In order to understand this point, we discuss the effect of $\text{SL}_2(\mathbb{Z})$-transformations on the periods.

We present a compact formula for the two-loop sunrise integral with equal masses 
around two space-time dimensions, valid to all orders in dimensional regularisation parameter $\eps$.
This formula reads
\begin{align}
\label{main_result_intro}
 S_{111}\left(2-2\eps,t\right)
 &= 
 \frac{\psi_1(q_2)}{\pi}
 e^{-\eps I(f_2;q_2) -2\eps L -2\gamma_E \eps + 2\sum\limits_{n=2}^\infty \frac{\left(-1\right)^n}{n} \zeta_n \eps^n} 
 \nonumber \\
 &
 \left\{
 \left[
 \sum\limits_{j=0}^\infty 
 \left(
 \eps^{2j} I\left(\left\{1,f_4\right\}^j;q_2 \right)
 -
 \frac{1}{2} \eps^{2j+1} I\left(\left\{1,f_4\right\}^j,1;q_2 \right)
 \right)
 \right]
 \sum\limits_{k=0}^\infty \eps^k B^{(k)}\left(2,0\right)
 \right. \nonumber \\
 & 
 \left.
 +
  \sum\limits_{j=0}^\infty 
   \eps^j 
   \sum\limits_{k=0}^{\lfloor \frac{j}{2} \rfloor} I\left( \left\{1,f_4\right\}^k, 1, f_3, \left\{f_2\right\}^{j-2k}; q_2\right)
 \right\}.
\end{align}
where the full $\eps$-dependence on the right-hand side is explicit.
The notation is explained in detail in the main part of this paper, the essentials are as follows:
We denote by $I(f_{i_1},...,f_{i_n};q_2)$ an $n$-fold iterated integral, where the letters $f_{i_1}$, .., $f_{i_n}$ are
modular forms and $q_2$ the nome of the elliptic curve.
Repeated sequences of letters are abbreviated as in $\{f_{i_1},f_{i_2}\}^3 = f_{i_1},f_{i_2},f_{i_1},f_{i_2},f_{i_1},f_{i_2}$.
The function $\psi_1(q_2)$ is a period of the elliptic curve
(and a modular form of weight $1$ for the relevant congruence subgroup),
the variable $L$ equals $L=\ln(m^2/\mu^2)$, and the $B^{(k)}\left(2,0\right)$'s are boundary constants.
From eq.~(\ref{main_result_intro}) one easily obtains the $j$-th term of the $\eps$-expansion of $S_{111}(2-2\eps,t)$
by expanding the exponential function and by collecting all terms proportional to $\eps^j$.
Products of iterated integrals may be converted to a linear combination of single iterated integrals
with the help of the shuffle product.
Eq.~(\ref{main_result_intro}) is not only helpful in retrieving quickly the $j$-th term of the $\eps$-expansion,
it will also be useful to elaborate and to extend the recently proposed 
coaction for Feynman integrals \cite{Brown:2015aa,Abreu:2017enx,Abreu:2017mtm}.

Our results are based on comparing $q$-series expansions to high orders.
A typical value for the order of the expansion is ${\mathcal O}(q^{1000})$.
It is clear that for a pure mathematician this does not constitute a strict proof.
However, the use of ``experimental mathematics'', like for example the PSLQ-algorithm \cite{Ferguson:1992},
is standard practice in theoretical particle physics.
The fact, that our equations agree to a very high power in the $q$-expansion gives us sufficient confidence that the equations are correct.
Let us mention that it is not too difficult to prove our results in a strict mathematical sense.
The specific elliptic curve associated to the Feynman parameter representation 
has been studied in the mathematical literature \cite{Beauville:1982,Stienstra:1985,Sebbar:2002aa,Maier:2006aa}.
Using the existing results in the mathematical literature, the fact that the spaces of modular forms are finite-dimensional,
and the first few terms in the $q$-expansion will constitute a strict mathematical proof.
However, the focus of our paper is on generic methods. Thus we will not make use of existing results on specific elliptic curves.

This paper is organised as follows:
\ifthenelse{\boolean{arxiveversion}}
{
In section~\ref{sec:IntroModEisEich} we recall basic definitions and facts about modular forms.
This section will be familiar for mathematicians, however since we address an audience consisting of 
mathematicians and physicists, we feel that this section will be helpful for physicists.
We review the definitions of modular forms for the full modular group and the standard congruence subgroups.
We introduce Dirichlet characters and generalised Eisenstein series.
We present a theorem due to Ligozat, which allows us to decide whether a quotient of eta functions is a modular form
or not.
Finally, we introduce in this section iterated integrals of modular forms and discuss the special case of
Eichler integrals.
} 
{
In section~\ref{sec:IntroModEisEich} we recall a few basic facts about modular forms,
we introduce iterated integrals of modular forms and discuss the special case of
Eichler integrals.
} 
Section~\ref{sec:Sunrise} introduces the two-loop sunrise integral.
We give the basic definitions and present the differential equation satisfied by the two-loop sunrise integral.
In section~\ref{sec:EllipticCurve} we discuss one elliptic curve associated to the two-loop sunrise integral,
This elliptic curve is obtained from the zero set of the second graph polynomial.
Section~\ref{sec:integration_kernels} is one of the main sections of this paper.
In this section we show that all integration kernels for the equal mass sunrise integral and the kite
integral can be expressed in terms of modular forms for the congruence subgroup $\Gamma_0(12)$ with characters.
It follows that these two Feynman integrals can be expressed to all orders in $\eps$ in terms of iterated integrals
of modular forms.
For the sunrise integral we present in this section 
the explicit results for the first three terms in the $\eps$-expansion.
The first term in the $\eps$-expansion of the sunrise integral is an Eichler integral and we
investigate in section~\ref{sec:TrafoSunrise} the transformation properties of this term
under modular transformations of $\Gamma_0(12)$.
In section~\ref{sec:all_order} we derive eq.~(\ref{main_result_intro}) by solving the appropriate differential
equation to all orders in $\eps$ (as opposed to order-by-order in $\eps$).
In section~\ref{sect:Gamma_6} we repeat the calculation with the elliptic curve obtained from the maximal cut of the sunrise integral.
With a standard choice of periods this gives us modular forms of level $6$.
In section~\ref{sect:choice_periods} we discuss the effect of the freedom to choose a different pair of periods spanning the lattice,
related to the original pair by a $\text{SL}_2(\mathbb{Z})$-transformation.
Finally, our conclusions are given in section~\ref{sect:conclusions}.
\ifthenelse{\boolean{arxiveversion}}
{
The appendix contains useful information on Kronecker symbols (appendix~\ref{sec:Kronecker_symbol}),
the definition of the $\mathrm{ELi}$-functions (appendix~\ref{sect:ELi})
and
a summary on all modular forms of level $12$ relevant to the sunrise integral and the kite integral (appendix~\ref{sect:summary_modular_forms}).
} 
{
The appendix contains useful information 
on generalised Eisenstein series (appendix~\ref{sect:eisenstein_space}),
on the definition of the $\mathrm{ELi}$-functions (appendix~\ref{sect:ELi})
and
a summary on all modular forms of level $12$ relevant to the sunrise integral and the kite integral (appendix~\ref{sect:summary_modular_forms}).
} 

\section{Review of basic definitions and facts}
\label{sec:IntroModEisEich}

\ifthenelse{\boolean{arxiveversion}}
{
General and detailed introductions to modular forms can be found 
in many textbooks \cite{Bruinier,Diamond,Kilford,Lang,Miyake,Shimura,Stein}. 
In the following we will review the main definitions and properties of modular forms, Eisenstein series, 
eta quotients, iterated integrals and Eichler integrals,
focusing on the aspects needed in the sequel of the article.
The first part of this short introduction follows the textbooks mentioned above.

\subsection{Modular forms}
Let the complex upper half plane be denoted by 
\begin{align}
\mathbb{H} = \lbrace \; \tau \in \mathbb{C} \; | \; \text{Im}(\tau) > 0 \; \rbrace.
\end{align}
The (full) modular group $\text{SL}_2(\mathbb{Z})$ is the group of $(2 \times 2)$-matrices over the integers with unit determinant:
\begin{align}
\text{SL}_2(\mathbb{Z}) = 
\left\{ 
\left( \begin{array}{cc}
a & b \\ 
c & d
\end{array}  \right)  
 \bigg\vert \
a,b,c,d \in \mathbb{Z},\ ad-bc=1
\right\}.
\end{align}
This group acts on elements $\tau$ of the Riemann sphere $\hat{\mathbb{C}} = \mathbb{C} \cup \lbrace \infty \rbrace$ by M\"obius transformations:
\begin{align}
 \gamma(\tau) = \dfrac{a\tau+b}{c\tau+d}, 
 \qquad 
 \gamma = 
\left( \begin{array}{cc}
a & b \\ 
c & d
\end{array}  \right) \in \text{SL}_2(\mathbb{Z}),
 \qquad 
 \tau \in \hat{\mathbb{C}},
\end{align}
transforming the complex upper half plane into itself.
The modular group is generated by the matrices
\begin{align}
T= \left( \begin{array}{rr}
1 & 1 \\ 
0 & 1
\end{array}  \right) \qquad \text{and} \qquad S= \left( \begin{array}{rr}
0 & -1 \\ 
1 & 0
\end{array}  \right),
\end{align}
where the '$T$' stands for ``translation'', describing the action of the matrix $T$ on $\hat{\mathbb{C}}$, given by $T(\tau)=\tau+1$.
\begin{Def}
Let $k$ be an integer. A meromorphic function $f: \mathbb{H} \rightarrow \mathbb{C}$ is weakly modular of weight $k$ 
for $\text{SL}_2(\mathbb{Z})$ if
\begin{align}
f(\gamma(\tau)) = f\left( \dfrac{a\tau+b}{c\tau+d} \right) = (c\tau+d)^k \cdot f(\tau) 
 \qquad \text{for} \;\; \gamma = \left( \begin{array}{cc}
a & b \\ 
c & d
\end{array} \right) \in \text{SL}_2(\mathbb{Z})
 \;\; \text{and} \;\; 
 \tau \in \mathbb{H}.
\end{align}
\end{Def}
It can be shown that $f$ is weakly modular of weight $k$ for $\text{SL}_2(\mathbb{Z})$ if
\begin{align}
f(T(\tau)) = f(\tau+1) = f(\tau) \qquad \text{and} \qquad f(S(\tau)) = f(-1/\tau) = \tau^k f(\tau). 
\label{fTS}
\end{align}
Weak modularity is one of the three conditions a function $f$ has to fulfil to be called \textit{modular form}:
\begin{Def}
Let $k$ be an integer. A meromorphic function $f: \mathbb{H} \rightarrow \mathbb{C}$ is a modular form of weight $k$ for $\text{SL}_2(\mathbb{Z}$) if
\begin{itemize}
\item[(i)] $f$ is weakly modular of weight $k$ for $\text{SL}_2(\mathbb{Z})$,
\item[(ii)] $f$ is holomorphic on $\mathbb{H}$,
\item[(iii)] $f$ is holomorphic at $\infty$.
\end{itemize}
The set of modular forms of weight $k$ for $SL_2(\mathbb{Z})$ is denoted by $\mathcal{M}_k(\text{SL}_2(\mathbb{Z}))$. 
\end{Def}
To motivate $(iii)$ we recall that $f$ is $\mathbb{Z}$-periodic, as we saw in eq. (\ref{fTS}).
It follows that $f$ has a Fourier expansion
\begin{align}
f(\tau) = \sum\limits_{n \in \mathbb{Z}} a_n q^n \qquad \text{with} \qquad q=e^{2\pi i \tau}.
\end{align}
With the relation $|q|=e^{-2\pi \text{Im}(\tau)}$ we find that $q \rightarrow 0$ for $\text{Im}(\tau) \rightarrow \infty$. 
Thus, we can define $f$ to be holomorphic at $\infty$ if the Fourier expansion is holomorphic at $q=0$, 
i.e. $a_n=0$ for $n<0$, ending up with
\begin{align}
f(\tau) = \sum\limits_{n=0}^{\infty} a_n q^n \qquad \text{with} \qquad q=e^{2\pi i \tau}.
\end{align}
This Fourier expansion is called the \textit{$q$-expansion of the function f}.\\
We define the value at $\infty$ of a modular form $f$ for $\text{SL}_2(\mathbb{Z})$ as 
\begin{align}
f(\infty) = a_0.
\end{align}
The term $a_0$ of the $q$-expansion plays an important role for distinguishing modular forms:
\begin{Def}
Let $f$ be a modular form of weight $k$ for $\text{SL}_2(\mathbb{Z})$. If 
\begin{align}
a_0=0
\end{align}
in the Fourier expansion of $f$, i.e.
\begin{align}
f(\tau) = \sum\limits_{n=1}^{\infty} a_n q^n \qquad \text{with} \qquad q=e^{2\pi i \tau},
\end{align}
then $f$ is called a cusp form of weight $k$ for $\text{SL}_2(\mathbb{Z})$. 
The set of cusp forms of weight $k$ for $\text{SL}_2(\mathbb{Z})$ is denoted by $\mathcal{S}_k(\text{SL}_2(\mathbb{Z}))$.
\end{Def}
Let us look at products of modular forms.
Assume that $f$ and $g$ are modular forms for $\text{SL}_2(\mathbb{Z})$
of weight $k$ and $l$, respectively.
Then the function $f \cdot g$ is a modular form for $\text{SL}_2(\mathbb{Z})$
with weight $(k+l)$.

\subsubsection{Modular forms for congruence subgroups}

We may extend the considerations above towards subgroups of $\text{SL}_2(\mathbb{Z})$. 
The \textit{standard congruence subgroups} of the modular group $\text{SL}_2(\mathbb{Z})$ are defined by
\begin{align}
\Gamma_0(N) &= \left\{ \left( \begin{array}{cc}
a & b \\ 
c & d
\end{array}  \right) \in \text{SL}_2(\mathbb{Z}): c \equiv 0\ \text{mod}\ N \right\}, 
 \nonumber \\
\Gamma_1(N) &= \left\{ \left( \begin{array}{cc}
a & b \\ 
c & d
\end{array}  \right) \in \text{SL}_2(\mathbb{Z}): a,d \equiv 1\ \text{mod}\ N, \; c \equiv 0\ \text{mod}\ N  \right\},
 \nonumber \\
\Gamma(N) &= \left\{ \left( \begin{array}{cc}
a & b \\ 
c & d
\end{array}  \right) \in \text{SL}_2(\mathbb{Z}): a,d \equiv 1\ \text{mod}\ N, \; b,c \equiv 0\ \text{mod}\ N \right\}.
\end{align}
The group $\Gamma(N)$ is called the \textit{principal congruence subgroup}.
We have the inclusions
\begin{align}
\Gamma(N) \subseteq \Gamma_1(N) \subseteq \Gamma_0(N) \subseteq \text{SL}_2(\mathbb{Z}).
\end{align}
Let us also introduce the 
\textit{weight $k$ operator} $\slashoperator{\gamma}{k}$ acting on functions $f$ from $\mathbb{H}$ to $\mathbb{C}$:
\begin{Def}
Let $k$ be an integer, $f$ a function $f: \mathbb{H} \rightarrow \mathbb{C}$ and $\gamma \in \text{SL}_2(\mathbb{Z})$.
The action of the weight $k$ operator $\slashoperator{\gamma}{k}$ on $f$ is given by
\begin{align}
(f \slashoperator{\gamma}{k})(\tau) = (c\tau+d)^{-k} \cdot f(\gamma(\tau)).
\end{align}
\end{Def}
With this definition we may re-write the condition for a meromorphic function
$f: \mathbb{H} \rightarrow \mathbb{C}$ to be weakly modular of weight $k$ 
for $\text{SL}_2(\mathbb{Z})$ as
\begin{align}
 f \slashoperator{\gamma}{k} = f
 \qquad \text{for all} \;\; \gamma \in \text{SL}_2(\mathbb{Z}).
\end{align}
To define a modular form $f$ for a congruence subgroup $\Gamma$ of $\text{SL}_2(\mathbb{Z})$ 
we adopt the weak modularity of $f$ with respect to $\Gamma$ and extend the holomorphy condition. 
A meromorphic function $f: \mathbb{H} \rightarrow \mathbb{C}$ is weakly modular of weight $k$ 
for $\Gamma$ if
\begin{align}
 f \slashoperator{\gamma}{k} = f
 \qquad \text{for all} \;\; \gamma \in \Gamma.
\end{align}
For each congruence subgroup $\Gamma$ of $\text{SL}_2(\mathbb{Z})$ there is a smallest positive integer
$N$, such that $\Gamma(N) \subseteq \Gamma$.
It follows that $\Gamma$ contains a translation matrix $T_{\Gamma}$ given by 
\begin{align}
T_{\Gamma} = \left( \begin{array}{cc}
1 & N \\ 
0 & 1
\end{array}  \right),\qquad \text{with} \;\; T_{\Gamma}(\tau)=\tau+N.
\end{align}
This implies that a weakly modular function $f$ for $\Gamma$ is $N\mathbb{Z}$-periodic.
$N$ is called the \textit{level} of $f$.
If in addition $f$ is holomorphic at $\infty$, it has a Fourier expansion of the form
\begin{align}
f(\tau) = \sum\limits_{n=0}^{\infty} a_n q^n_N \qquad \text{with} \qquad q_N = e^{2\pi i \tau/N}.
\end{align} 
Modular forms for $\Gamma$ need not only be holomorphic on $\mathbb{H}$ and $\infty$, but also
on rational points on the real axis.
Let us denote by $\overline{\mathbb{H}}$ the extended upper half plane:
\begin{align}
\overline{\mathbb{H}} = \mathbb{H} \cup \lbrace \infty \rbrace \cup \mathbb{Q}.
\end{align} 
The action of $\Gamma$ divides the set $\lbrace \infty \rbrace \cup \mathbb{Q}$ into equivalence classes 
with respect to $\Gamma$. 
Such an equivalence class (or -- by an abuse of notation -- a rational number or infinity representing an equivalence class) is called a \textit{cusp of $\Gamma$}. 
Every $s \in \mathbb{Q}$ can be expressed as $s=\alpha(\infty)$ 
for some $\alpha \in \text{SL}_2(\mathbb{Z})$, 
which enables us to define the holomorphy at the cusps with the help of the weight $k$ operator.
This leads to the following definition of modular forms for $\Gamma$:
\begin{Def}
Let $\Gamma$ be a congruence subgroup of $\text{SL}_2(\mathbb{Z})$ and let $k$ be an integer. A meromorphic function $f: \mathbb{H} \rightarrow \mathbb{C}$ is a modular form of weight $k$ for $\Gamma$ if
\begin{itemize}
\item[(i)] $f$ is weakly modular of weight $k$ for $\Gamma$,
\item[(ii)] $f$ is holomorphic on $\mathbb{H}$,
\item[(iii)] $f \slashoperator{\alpha}{k}$ is holomorphic at $\infty$ for all $\alpha \in \text{SL}_2(\mathbb{Z})$.
\end{itemize} 
If, additionally,
\begin{itemize}
\item[(iv)] $a_0=0$ in the Fourier expansion of $f \slashoperator{\alpha}{k}$ for all $\alpha \in \text{SL}_2(\mathbb{Z})$,
\end{itemize}
we say that $f$ is a cusp form of weight $k$ with respect to $\Gamma$. The space of modular forms of weight $k$ for $\Gamma$ is denoted by $\mathcal{M}_k(\Gamma)$, the space of cusp forms of weight $k$ for $\Gamma$ by $\mathcal{S}_k(\Gamma)$. 
\end{Def}
The space $\mathcal{M}_k(\Gamma)$ is a finite dimensional $\mathbb{C}$-vector space.
Furthermore, $\mathcal{M}_k(\Gamma)$ is the direct sum of two finite dimensional $\mathbb{C}$-vector spaces:
the space of cusp forms $ \mathcal{S}_k(\Gamma)$ and the \textit{Eisenstein subspace} $\mathcal{E}_k(\Gamma)$, 
to be defined in section~\ref{sect:eisenstein_space}.
We therefore have
\begin{align}
\dim (\mathcal{M}_k(\Gamma)) = \dim(\mathcal{S}_k(\Gamma)) + \dim(\mathcal{E}_k(\Gamma)).
\end{align}
We have the inclusions
\begin{align}
 \mathcal{M}_k(\text{SL}_2(\mathbb{Z})) \subseteq \mathcal{M}_k(\Gamma_0(N)) \subseteq \mathcal{M}_k(\Gamma_1(N)) \subseteq \mathcal{M}_k(\Gamma(N))
\end{align}
and
\begin{align}
 \mathcal{S}_k(\text{SL}_2(\mathbb{Z})) \subseteq \mathcal{S}_k(\Gamma_0(N)) \subseteq \mathcal{S}_k(\Gamma_1(N)) \subseteq \mathcal{S}_k(\Gamma(N)).
\end{align}

\subsubsection{Dirichlet characters}

The group $\Gamma_1(N)$ is a normal subgroup of $\Gamma_0(N)$ and there is a group isomorphism
\begin{align}
 \Gamma_0(N)/\Gamma_1(N) \sim (\mathbb{Z}/N\mathbb{Z})^{\times},
\end{align}
where $(\mathbb{Z}/N\mathbb{Z})^{\times}$ denotes the multiplicative group of units.
This isomorphism is given by
\begin{align}
\left( \begin{array}{cc}
a & b \\ 
c & d
\end{array}  \right)  
 \rightarrow 
 d \ \text{mod} \ N.
\end{align}
This motivates the definition of a modular form with a \textit{Dirichlet character}. 
We first recall the definition of a Dirichlet character:
\begin{Def}
Let $N$ be a positive integer. A Dirichlet character modulo $N$
is a function $\chi: (\mathbb{Z}/N\mathbb{Z})^{\times} \rightarrow \mathbb{C}^{\times}$ that is a homomorphism of groups,
i.e. $\chi(nm) = \chi(n) \chi(m)$ for all $n,m \in (\mathbb{Z}/N\mathbb{Z})^{\times}$.
We may extend $\chi$ to a function $\chi: \mathbb{Z}/N\mathbb{Z} \rightarrow \mathbb{C}$ by setting
$\chi(n)=0$ if $\gcd(n,N) > 1$ and then further extend to a function $\chi: \mathbb{Z} \rightarrow \mathbb{C}$
by setting $\chi(n) = \chi(n \mod N)$.
By abuse of notation we denote both extensions again by $\chi$.
This function satisfies
\begin{itemize}
\item[(i)] $\chi(n) = \chi(n+N) \qquad \forall n \in \mathbb{Z}$,
\item[(ii)] $\chi(n)=0$ \ if \ $\gcd(n,N) > 1$ \quad and \quad $\chi(n) \neq 0$ \ if \ $\gcd(n,N) =1$,
\item[(iii)] $\chi(nm) = \chi(n) \chi(m) \qquad \forall\ n,m \in \mathbb{Z}$. 
\end{itemize}
We denote by $\chi_0$ the trivial character modulo $N$ 
(with $\chi_0(n) = 1$ if $\gcd(n,N)=1$ and $\chi_0(n)=0$ otherwise).
The trivial character modulo $1$ is denoted by $1$.
The conductor of $\chi$ is the smallest positive divisor $d|N$ such that there is a character $\chi'$ modulo $d$
with
\begin{align}
 \chi(n) = \chi'(n) \qquad \forall\ n \in \mathbb{Z} \qquad \text{with}\ \gcd(n,N) = 1.
\end{align}
A Dirichlet character is called primitive, if its modulus equals its conductor.
\end{Def}
Property $(ii)$ and $(iii)$ imply:
\begin{align}
\chi(1)=1.
\end{align}
If $\chi$ is a Dirichlet character modulo $N$ and $M$ a positive integer, 
$\chi$ induces a Dirichlet character $\tilde{\chi}$ with modulus $(M \cdot N)$ by setting
\begin{align}
 \tilde{\chi}(n)
 =
 \left\{
 \begin{array}{rl}
  \chi(n), & \mbox{if} \; \mbox{gcd}(n,M \cdot N) = 1, \\
  0, & \mbox{if} \; \mbox{gcd}(n,M \cdot N) \neq 1. \\
 \end{array}
 \right.
\end{align}
In the other direction we may associate to a Dirichlet character $\chi$ with modulus $N$ and conductor $d$
a primitive Dirichlet character $\bar{\chi}$ modulo $d$ as follows:
We first note if $\mbox{gcd}(n,d)=1$ there exists an integer $n'$ such that $\mbox{gcd}(n',N)=1$ and $n' \equiv n \mod d$.
We set
\begin{align}
 \bar{\chi}(n)
 =
 \left\{
 \begin{array}{rl}
  \chi(n'), & \mbox{if} \; \mbox{gcd}(n,d) = 1, \\
  0, & \mbox{if} \; \mbox{gcd}(n,d) \neq 1. \\
 \end{array}
 \right.
\end{align}
The character of modulus $N$ induced by the primitive character $\bar{\chi}$ is again $\chi$.

With the help of a Dirichlet character we may generalise the transformation behaviour of modular forms
and define a modular form with character $\chi$:
\begin{Def}
Let $N$ be a positive integer and let $\chi$ be a Dirichlet character modulo $N$. A function $f: \mathbb{H} \rightarrow \mathbb{C}$ is a modular form of weight $k$ for $\Gamma_0(N)$ with character $\chi$ if
\begin{itemize}
\item[(i)] $f$ is holomorphic on $\mathbb{H}$,
\item[(ii)] $f$ is holomorphic at the cusps of $\Gamma_1(N)$,
\item[(iii)] $f \left( \dfrac{a\tau+b}{c\tau+d}\right) = \chi(d) (c\tau+d)^k f(\tau)$ $\qquad \text{for} \;\; \left( \begin{array}{cc}
a & b \\ 
c & d
\end{array}  \right) \in \Gamma_0(N)$. 
\end{itemize}
The space of modular forms of weight $k$ and character $\chi$ for the congruence subgroup $\Gamma_0(N)$ is denoted by $\mathcal{M}_k(N,\chi)$ and the associated space of cusp forms by $\mathcal{S}_k(N,\chi)$.
\end{Def}
It can be shown that the space $\mathcal{M}_k(\Gamma_1(N))$ is a direct sum of spaces of modular forms with characters:
\begin{align}
\label{mod_form_space_decomp}
\mathcal{M}_{k}(\Gamma_1(N)) = \bigoplus\limits_{\chi}\ \mathcal{M}_k(N,\chi),
\end{align}
where the sum runs over all Dirichlet characters modulo $N$.
We have similar decompositions for the space of cusp forms and the Eisenstein subspaces:
\begin{align}
\mathcal{S}_{k}(\Gamma_1(N)) = \bigoplus\limits_{\chi}\ \mathcal{S}_k(N,\chi),
 \qquad
\mathcal{E}_{k}(\Gamma_1(N)) = \bigoplus\limits_{\chi}\ \mathcal{E}_k(N,\chi).
\end{align}

\subsubsection{Eisenstein series}
\label{sect:eisenstein_space}

A basis for the Eisenstein subspace $\mathcal{E}_k(N,\chi)$ can be given explicitly.
To this aim we first define \textit{generalised Eisenstein series}.
Let $\phi$ and $\psi$ be primitive Dirichlet characters with conductors $L$ and $M$, respectively.
We set
\begin{align}
E_{k} (\tau; \phi,\psi) &
 = 
 a_0 + \sum\limits_{m=1}^{\infty} \left( \sum\limits_{d|m} \psi(d) \cdot \phi(m/d) \cdot d^{k-1} \right) q_M^{m},
 \qquad q_M = e^{2\pi i \tau/M}.
\end{align}
The normalisation is such that the coefficient of $q_M$ is one.
The constant term $a_0$ is given by
\begin{align}
 a_0 &
 = 
 \begin{cases}
 -\frac{B_{k,\psi}}{2k}, \qquad &\text{if}\ L=1, \\
0, \qquad &\text{if}\ L>1.
\end{cases}
\end{align}
Note that $L$ denotes the conductor of $\phi$ and the constant term $a_0$ depends therefore on $\phi$ and $\psi$.
The generalised Bernoulli-numbers $B_{k,\psi}$ are defined by
\begin{align}
 \sum\limits_{m=1}^{M} \psi(m) \dfrac{xe^{mx}}{e^{Mx}-1}
 =
 \sum\limits_{k=0}^{\infty} B_{k,\psi} \dfrac{x^k}{k!}. 
\end{align}
The generalised Eisenstein series are modular forms \cite{Stein}:
\begin{Theo}
\label{Eisenstein_series_are_modular}
Suppose $K$ is a positive integer, the Dirichlet characters $\phi$, $\psi$ are as above and $k$ is a positive integer such that
$\phi(-1) \psi(-1) = (-1)^k$.
For $k=1$ we require in addition $\phi(-1)=1$ and $\psi(-1)=-1$.
Except when $k=2$ and $\phi=\psi=1$, the Eisenstein series $E_k(K \tau;\phi,\psi)$ defines an element of
$\mathcal{M}_k(KLM,\tilde{\chi})$, where $\tilde{\chi}$ is the Dirichlet character with modulus $KLM$ induced by $\phi\psi$.
In the case $k=2$, $\phi=\psi=1$ and $K>1$ we use the notation $E_2(\tau)=E_2(\tau;\phi,\psi)$ 
and $B_{2,K}(\tau) = E_2(\tau)- K E_2(K \tau)$.
Then $B_{2,K}(\tau)$ is a modular form in $\mathcal{M}_2(\Gamma_0(K))$.
\end{Theo}
We may now give a basis for the Eisenstein subspace $\mathcal{E}_k(N,\chi)$:
\begin{Theo}
The Eisenstein series in $\mathcal{M}_k(N,\chi)$ coming from theorem~\ref{Eisenstein_series_are_modular} with $KLM | N$ and $\chi$ the Dirichlet character of modulus $N$ induced from $\phi\psi$ form 
a basis for the Eisenstein subspace $\mathcal{E}_k(N,\chi)$.
\end{Theo}

\subsubsection{An example}

Let us investigate as an example modular forms of weight $1$ and $2$ for $\Gamma_1(12)$.
The choice of this example is motivated by the fact, that these modular forms will show up again in the rest of the paper.
At level $N=12$ we have $4$ Dirichlet characters.
We denote these characters by $\chi_0$, $\chi_1$, $\chi_2$ and $\chi_3$.
These characters are induced by primitive characters $\bar{\chi}_0$, $\bar{\chi}_1$, $\bar{\chi}_2$ and $\bar{\chi}_3$.
The latter are given in terms of the Kronecker symbol by
\begin{align}
\bar{\chi}_0 = \left( \dfrac{1}{n} \right), \qquad
\bar{\chi}_1 = \left( \dfrac{-3}{n} \right), \qquad
\bar{\chi}_2 = \left( \dfrac{-4}{n} \right), \qquad
\bar{\chi}_3 = \left( \dfrac{12}{n} \right).
\end{align}
The definition of the \textit{Kronecker symbol} is given in appendix~\ref{sec:Kronecker_symbol}.
$\chi_0$ denotes the trivial character.
The conductors of the characters $\chi_0$, $\chi_1$, $\chi_2$ and $\chi_3$ are
$1$, $3$, $4$ and $12$, respectively.
At weight $1$ the space of modular forms decomposes as
\begin{align}
 \mathcal{M}_{1}(\Gamma_1(12)) = 
 \mathcal{M}_1(12,\chi_1) \oplus \mathcal{M}_1(12,\chi_2),
\end{align}
the spaces $\mathcal{M}_1(12,\chi_0)$ and $\mathcal{M}_1(12,\chi_3)$ are empty.
There are no cusp forms
\begin{align}
 \dim(\mathcal{S}_1(12,\chi_1)) = \dim(\mathcal{S}_1(12,\chi_2)) = 0,
\end{align}
and we have therefore
\begin{align}
 \mathcal{M}_{1}(\Gamma_1(12)) = 
 \mathcal{E}_1(12,\chi_1) \oplus \mathcal{E}_1(12,\chi_2).
\end{align}
A basis for the Eisenstein spaces is given by
\begin{align}
 & \text{Basis of}\ \mathcal{E}_1(12,\chi_1) = \lbrace E_1(\tau;\bar{\chi}_0,\bar{\chi}_1), E_1(2\tau;\bar{\chi}_0,\bar{\chi}_1), E_1(4\tau;\bar{\chi}_0,\bar{\chi}_1) \rbrace, \nonumber \\
 & \text{Basis of}\ \mathcal{E}_1(12,\chi_2) = \lbrace E_1(\tau;\bar{\chi}_0,\bar{\chi}_2), E_1(3\tau;\bar{\chi}_0,\bar{\chi}_2) \rbrace.
\end{align}
At weight $2$ we have the decomposition
\begin{align}
 \mathcal{M}_{2}(\Gamma_1(12)) = 
 \mathcal{M}_2(12,\chi_0) \oplus \mathcal{M}_2(12,\chi_3),
\end{align}
the spaces $\mathcal{M}_2(12,\chi_1)$ and $\mathcal{M}_2(12,\chi_2)$ are empty.
Again, there are no cusp forms
\begin{align}
 \dim(\mathcal{S}_2(12,\chi_0)) = \dim(\mathcal{S}_2(12,\chi_3)) = 0,
\end{align}
and we have therefore
\begin{align}
 \mathcal{M}_{2}(\Gamma_1(12)) = 
 \mathcal{E}_2(12,\chi_0) \oplus \mathcal{E}_2(12,\chi_3).
\end{align}
A basis for the Eisenstein spaces is given by
\begin{align}
 & \text{Basis of}\ \mathcal{E}_2(12,\chi_0) 
 = \left\{ B_{2,2}(\tau), B_{2,3}(\tau), B_{2,4}(\tau), B_{2,6}(\tau), B_{2,12}(\tau) \right\}, \nonumber \\
 & \text{Basis of}\ \mathcal{E}_2(12,\chi_3) = \lbrace E_2(\tau;\bar{\chi}_0,\bar{\chi}_3), E_2(\tau;\bar{\chi}_1,\bar{\chi}_2), E_2(\tau;\bar{\chi}_2,\bar{\chi}_1), E_2(\tau;\bar{\chi}_3,\bar{\chi}_0) \rbrace.
\end{align}

\subsection{Eta quotients}
\label{sect_eta_quotients}

\textit{Dedekind's eta function} is defined by
\begin{align}
 \eta (\tau) = e^{\frac{i \pi \tau}{12}} \prod\limits_{n=1}^{\infty} (1-e^{2\pi i n \tau}) 
 =
 q^{\frac{1}{24}} \prod\limits_{n=1}^{\infty} (1-q^n),
 \qquad q=e^{2\pi i \tau}.
\end{align}
Very often we encounter quotients of eta functions, which we may write as
\begin{align}
 Q(\tau) = \prod\limits_{1 \leq d | N} \eta (d\tau)^{r_d},
\end{align}
with $r_d \in \mathbb{Z}$.
We would like to know under which conditions the function $Q(\tau)$ is a modular form.
A criterion for an eta quotient to be a modular form has been given by Ligozat \cite{Ligozat:1975,Alaca:2015aa}.
\begin{Theo}
Let $Q(\tau)$ be an eta quotient which satisfies the following conditions:
\begin{itemize}
\item[(i)] $\sum\limits_{1 \leq d|N} d \cdot r_d \equiv 0 \mod  24$,
\item[(ii)] $\sum\limits_{1 \leq d|N} \frac{N}{d} \cdot r_d \equiv 0 \mod 24$,
\item[(iii)] For each $d'|N$, $\sum\limits_{1 \leq d|N} \frac{\gcd(d',d)^2 \cdot r_d}{d} \geq 0$.
\end{itemize}
Then $Q(\tau) \in \mathcal{M}_k(N,\chi)$, i.e. $Q(\tau)$ is a modular form of weight $k$ and level $N$ with character $\chi$. 
The weight is given by
\begin{align}
k = \dfrac{1}{2} \sum\limits_{1 \leq d|N} r_d.
\end{align}
Furthermore, if the number $s$ defined by
\begin{align}
 s = \prod\limits_{1 \leq d|N} d^{r_d}
\end{align}
is an integer, then the Dirichlet character $\chi$ is given by the Kronecker symbol
\begin{align}
\chi(n) = \left( \frac{(-1)^k s}{n}\right).
\label{CharacEtaQuo}
\end{align}
\label{TheoEtaMod}
\end{Theo}
Condition (iii) ensures that $Q(\tau)$ is holomorphic \cite{Koehler}.
Note that in practice the number $s$ can be rather large or not even an integer.
On the other hand there are only a finite number of Dirichlet characters for a given level $N$.
It is therefore simpler to start from the decomposition in eq.~(\ref{mod_form_space_decomp}) and to check to which space
$\mathcal{M}_k(N,\chi)$ the modular form $Q(\tau)$ belongs.
} 
{
We expect the reader to be familiar with modular forms.
General and detailed introductions to modular forms can be found 
in many textbooks \cite{Bruinier,Diamond,Kilford,Lang,Miyake,Shimura,Stein}. 
In addition, the arXiv version of this article contains a concise introduction to modular forms.
} 

\subsection{Iterated integrals of modular forms}
\label{sect:iterated_integrals_modular_forms}

Let $f_1(\tau)$, $f_2(\tau)$, ..., $f_n(\tau)$ be a set of modular form for a congruence subgroup $\Gamma$.
We denote the modular weight of the modular form $f_i(\tau)$ by $k_i$, in particular we do not require that all modular forms in the set have the same weight.
We define the $n$-fold \textit{iterated integral} \cite{Chen,Manin:2005,Hain:2014,Brown:2014aa} of these modular forms by
\begin{align}
I\left(f_1,f_2,...,f_n;\tau,\tau_0\right)
 & =
 \left(2 \pi i \right)^n
 \int\limits_{\tau_0}^{\tau} d\tau_1
 \int\limits_{\tau_0}^{\tau_1} d\tau_2
 ...
 \int\limits_{\tau_0}^{\tau_{n-1}} d\tau_n
 \;
 f_1\left(\tau_1\right)
 f_2\left(\tau_2\right)
 ...
 f_n\left(\tau_n\right).
\end{align}
With $q=\exp(2\pi i \tau)$ we may equally well write
\begin{align}
I\left(f_1,f_2,...,f_n;\tau,\tau_0\right)
 & =
 \int\limits_{q_0}^{q} \frac{dq_1}{q_1}
 \int\limits_{q_0}^{q_1} \frac{dq_2}{q_2}
 ...
 \int\limits_{q_0}^{q_{n-1}} \frac{dq_n}{q_n}
 \;
 f_1\left(\tau_1\right)
 f_2\left(\tau_2\right)
 ...
 f_n\left(\tau_n\right),
 \qquad \tau_j = \frac{1}{2\pi i} \ln q_j.
\end{align}
The definition includes the special case, where the first $(n-1)$ modular forms are the constant function $1$:
\begin{align}
\label{special_case_1_modular_form}
I(\underbrace{1,...,1}_{n-1},f_n;\tau,\tau_0)
 & =
 \int\limits_{q_0}^{q} \frac{dq_1}{q_1}
 \int\limits_{q_0}^{q_1} \frac{dq_2}{q_2}
 ...
 \int\limits_{q_0}^{q_{n-1}} \frac{dq_n}{q_n}
 \;
 f_n\left(\tau_n\right).
\end{align}
It will be convenient to introduce a short-hand notation for repeated letters.
We use the notation
\begin{align}
 \left\{f_i\right\}^j & =
 \underbrace{ f_i, f_i, ..., f_i}_{j}
\end{align}
to denote a sequence of $j$ letters $f_i$.
More generally we will use in the sequel the notation
\begin{align}
 \left\{ f_{i_1}, f_{i_2}, ..., f_{i_n} \right\}^j
 & =
 \underbrace{f_{i_1}, f_{i_2}, ..., f_{i_n}, ......, f_{i_1}, f_{i_2}, ..., f_{i_n}}_{j \;\; \mathrm{copies} \; \mathrm{of} \; f_{i_1}, f_{i_2}, ..., f_{i_n}}
\end{align}
to denote a sequence of $(j \cdot n)$ letters, consisting of $j$ copies of $f_{i_1}, f_{i_2}, ..., f_{i_n}$.
For example
\begin{align}
 \left\{ f_1, f_2 \right\}^3
 & =
 f_1, f_2, f_1, f_2, f_1, f_2. 
\end{align}
Thus we may write the left-hand side of eq.~(\ref{special_case_1_modular_form}) as 
\begin{align}
I(\{1\}^{n-1},f_n;\tau,\tau_0)
 & =
I(\underbrace{1,...,1}_{n-1},f_n;\tau,\tau_0).
\end{align}
We follow standard practice and define the zero-fold iterated integral to be one:
\begin{align}
I\left(;\tau,\tau_0\right)
 & = 1.
\end{align}
In analogy with the case of multiple polylogarithms we define the \textit{depth} of an iterated integral
$I(f_1,...,f_n;\tau,\tau_0)$ to be the number of iterated integrations $n$.
Note that in the case of multiple polylogarithms the depth is often called transcendental weight.
Here, it is more appropriate to use the word ``depth''. (Also the word ``length'' is used \cite{Brown:2014aa}).
The depth should not be confused with the modular weight, the former is the depth of the iterated integral,
the latter is associated to individual modular forms $f_1$, ..., $f_n$.
We have the shuffle product for iterated integrals
\begin{align}
 I\left(f_1,...,f_r;\tau,\tau_0\right)
 \cdot
 I\left(f_{r+1},...,f_r;\tau,\tau_0\right)
 & =
 \sum\limits_{\mathrm{shuffles} \; \sigma}
 I\left(f_{\sigma_1},...,f_{\sigma_n};\tau,\tau_0\right),
\end{align}
where the sum runs over all shuffles of $(1,...,r)$ and $(r+1,...,n)$, i.e. all permutations of $(1,...,n)$, which
keep the relative order of $(1,...,r)$ and $(r+1,...,n)$ fixed.

Our standard choice for the base point $\tau_0$ will be $\tau_0 = i \infty$, corresponding to $q_0=0$.
This is unproblematic for cusp forms. Here we have for a single integration
\begin{align}
 f = \sum\limits_{j=1}^\infty a_j q^j
 \qquad \Rightarrow \qquad
 \int\limits_0^q \frac{dq_1}{q_1} f = \sum\limits_{j=1}^\infty \frac{a_j}{j} q^j.
\end{align}
For non-cusp forms we proceed as follows: We first take $q_0$ to have a small non-zero value.
The integration will produce terms with $\ln(q_0)$. Let $R$ be the operator, which removes all $\ln(q_0)$-terms.
After these terms have been removed, we may take the limit $q_0\rightarrow 0$.
With a slight abuse of notation we set
\begin{align}
I\left(f_1,f_2,...,f_n;q\right)
 & =
 \lim\limits_{q_0\rightarrow 0}
 R \left[
 \int\limits_{q_0}^{q} \frac{dq_1}{q_1}
 \int\limits_{q_0}^{q_1} \frac{dq_2}{q_2}
 ...
 \int\limits_{q_0}^{q_{n-1}} \frac{dq_n}{q_n}
 \;
 f_1\left(\tau_1\right)
 f_2\left(\tau_2\right)
 ...
 f_n\left(\tau_n\right)
 \right].
\end{align}
This prescription is compatible with the shuffle product.
For a single integration we have for a non-cusp form
\begin{align}
 f = \sum\limits_{j=0}^\infty a_j q^j
 \qquad \Rightarrow \qquad
 I(f;q) = a_0 \ln(q) + \sum\limits_{j=1}^\infty \frac{a_j}{j} q^j.
\end{align}
The above prescription is familiar to physicists from regularisation and renormalisation.
For the more mathematical oriented reader let us mention that this prescription is equivalent to choosing a
tangential base point \cite{Brown:2014aa}.
This prescription is also commonly applied in the context of the integral representation of multiple polylogarithms with trailing zeros.

\subsection{Eichler integrals}

In general, an integral over a modular form does not have particular ``nice'' properties under modular transformations.
However, a $(k-1)$-fold integral over a modular form of weight $k$ does. 
These integrals are called \textit{Eichler integrals} \cite{Eichler:1957}.
In this paragraph we review Eichler integrals.

We have already seen that 
a modular form $f$ of weight $k$ and character $\chi$ transforms under a modular transformation $\gamma$ as
\begin{align}
\label{recap_trafo_modform_with_character}
 f(\gamma(\tau)) = \chi(d) (c\tau+d)^k f(\tau),
 \qquad 
 \gamma = \left( \begin{array}{cc} a & b \\ c & d \\ \end{array} \right).
\end{align}
Let $k \ge 2$. We may generalise the modular transformation law.
An Eichler integral $F$ of weight $(2-k)$ 
and character $\chi$ transforms under a modular transformation $\gamma$ as
\begin{align}
\label{trafo_eichler}
 F(\gamma(\tau)) = \chi(d) (c\tau+d)^{2-k} \left[  F(\tau) + P_\gamma(\tau) \right],
\end{align}
where $P_\gamma(\tau)$ is a polynomial in $\tau$ of degree at most $(k-2)$.
The polynomial $P_\gamma(\tau)$ is called the \textit{period polynomial}.
Let us denote
\begin{align}
 D = \frac{1}{2\pi i} \frac{d}{d\tau} = q \frac{d}{dq}.
\end{align}
If we set
\begin{align}
\label{modular_form_from_Eichler_integral}
 f(\tau) = D^{k-1} F(\tau),
\end{align}
it follows from Bol's identity \cite{Bol:1949}
\begin{align}
 \left(D^{k-1}F\right)\left(\gamma(\tau)\right)
 = \left(c\tau+d\right)^k D^{k-1}\left[ \left(c\tau+d\right)^{k-2} F\left(\gamma(\tau)\right) \right],
\end{align}
that $f(\tau)$ transforms under a modular transformation as in eq.~(\ref{recap_trafo_modform_with_character}).
Thus we obtain from an Eichler integral of weight $(2-k)$ a modular form of weight $k$ by
$(k-1)$-fold differentiation.

Let us now look in the reverse direction.
Given a modular form $f(\tau)$ of weight $k$ (and possibly with a character $\chi$) we set
\begin{align}
\label{def_Eichler_integral}
 F(\tau) =
 \frac{(2\pi i)^{k-1}}{(k-2)!}
 \int\limits_{\tau_0}^\tau d\sigma \; f(\sigma) \; (\tau-\sigma)^{k-2},
\end{align}
where $\tau_0$ denotes the fixed value of the lower integration boundary.
Then $F(\tau)$ transforms as in eq.~(\ref{trafo_eichler}), with the period polynomial given by
\begin{align}
 P_\gamma(\tau) =
 \frac{(2\pi i)^{k-1}}{(k-2)!}
 \int\limits_{\gamma^{-1}(\tau_0)}^{\tau_0} d\sigma \; f(\sigma) \; (\tau-\sigma)^{k-2}.
\end{align}
Thus, eq.~(\ref{def_Eichler_integral}) gives us an Eichler integral of the modular form $f(\tau)$.

Let us now investigate the set of all Eichler integrals corresponding to the modular
form $f(\tau)$.
Let $Q(\tau)$ be an arbitrary polynomial in $\tau$ of degree at most $(k-2)$.
If $F(\tau)$ is an Eichler integral (i.e. $F(\tau)$ transforms under modular transformations as in eq.~(\ref{trafo_eichler})), 
then
\begin{align}
 \tilde{F}(\tau) = F(\tau) + Q(\tau)
\end{align}
is an Eichler integral as well.
Given a modular form $f(\tau)$ of weight $k$ as above with Fourier expansion
\begin{align}
 f = \sum\limits_{n=0}^\infty a_n q^n
\end{align}
we see from eq.~(\ref{modular_form_from_Eichler_integral}) that all Eichler integrals of $f$ are of the form
\begin{align}
 F(\tau)
 =
 \frac{a_0}{(k-1)!} \left(\ln q \right)^{k-1}
 + \sum\limits_{n=1}^\infty \frac{a_n}{n^{k-1}} q^n
 + \sum\limits_{j=0}^{k-2} \frac{b_j}{j!} \left(\ln q \right)^j,
\end{align}
with arbitrary constants $b_0$, $b_1$, ..., $b_{k-2}$.
It will be convenient to set these arbitrary constants to zero and we will take
the Eichler integral of $f(\tau)$ to be
\begin{align}
\label{def_F_Eichler}
 F_{\mathrm{Eichler}}(\tau)
 =
 \frac{a_0}{(k-1)!} \left(\ln q \right)^{k-1}
 + \sum\limits_{n=1}^\infty \frac{a_n}{n^{k-1}} q^n,
\end{align}
unless specified otherwise.
In the notation of section~\ref{sect:iterated_integrals_modular_forms} this equals
\begin{align}
 F_{\mathrm{Eichler}}(\tau)
 =
 I(\{1\}^{k-2},f;q)
 =
 I(\underbrace{1,...,1}_{k-2},f;q).
\end{align}
We further have
\begin{align}
\label{F_Eichler}
 F_{\mathrm{Eichler}}(\tau)
 =
 \frac{(2\pi i)^{k-1}}{(k-2)!}
 \int\limits_{i \infty}^\tau d\sigma \; \left( f(\sigma) - a_0 \right) \; (\tau-\sigma)^{k-2}
 +
 \frac{a_0}{(k-1)!} \left(\ln q \right)^{k-1},
\end{align}
with $a_0=f(i \infty)$.

We note that one finds in the literature also slightly different definitions of an Eichler integral,
the definition given in \cite{Zagier:1991} is based on an analytic continuation of the $L$-series
and will give under the transformation of eq.~(\ref{trafo_eichler}) not a period polynomial but a period function.

\section{The two-loop sunrise integral}
\label{sec:Sunrise}

\begin{figure}
\begin{center}
\includegraphics[scale=1.0]{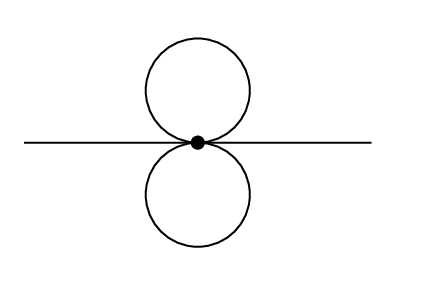}
\includegraphics[scale=1.0]{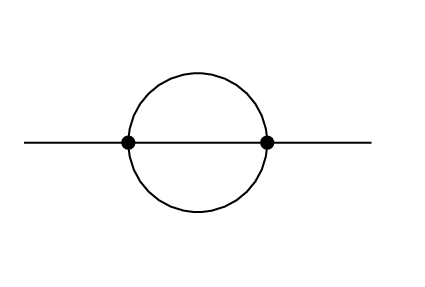}
\includegraphics[scale=1.0]{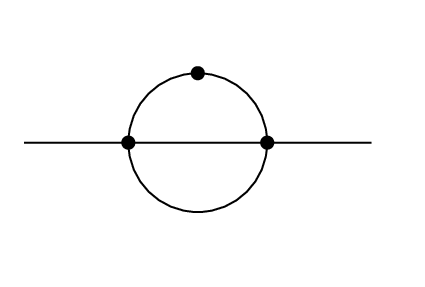}
\end{center}
\caption{
\it A set of master integrals for the sunrise family. 
A dot on a propagator indicates, that this propagator is raised to the power two.
}
\label{fig_master_integrals}
\end{figure}

We are interested in the family of Feynman integrals related to the two-loop sunrise graph with equal internal masses.
This family is given by
\begin{align}
\label{def_sunrise}
 S_{\nu_1\nu_2\nu_3}\left( D, p^2, m^2, \mu^2 \right)
 = 
 \left(\mu^2\right)^{\nu-D}
 \int \frac{d^Dk_1}{i \pi^{\frac{D}{2}}} \frac{d^Dk_2}{i \pi^{\frac{D}{2}}} \frac{d^Dk_3}{i \pi^{\frac{D}{2}}}
 \frac{\delta^D\left(p-k_1-k_2-k_3\right)}{\left(-k_1^2+m^2\right)^{\nu_1} \left(-k_2^2+m^2\right)^{\nu_2} \left(-k_3^2+m^2\right)^{\nu_3}},
\end{align}
with $\nu=\nu_1+\nu_2+\nu_3$ and $\nu_1,\nu_2,\nu_3 \in {\mathbb Z}$.
The dimension of space-time is denoted by $D$.
The arbitrary scale $\mu$ is introduced to keep the integral dimensionless.
The quantity $p^2$ denotes the momentum squared (with respect to the $D$-dimensional Minkowski metric defined by $g_{\mu\nu}=\mathrm{diag}(1, -1, -1, -1,...)$)
and we will write
\begin{align}
 t = p^2.
\end{align}
Where it is not essential we will suppress the dependence on the mass $m$ and the scale $\mu$ and simply write
$S_{\nu_1\nu_2\nu_3}( D, t)$ instead of $S_{\nu_1\nu_2\nu_3}( D, t, m^2, \mu^2)$.
In terms of Feynman parameters the two-loop integral is given by
\begin{align}
\label{def_Feynman_integral}
 S_{\nu_1\nu_2\nu_3}\left( D, t\right)
 =
 \frac{\Gamma\left(\nu-D\right)}{\Gamma(\nu_1)\Gamma(\nu_2)\Gamma(\nu_3)}
 \left(\mu^2\right)^{\nu-D}
 \int\limits_{\sigma} 
 x_1^{\nu_1-1} x_2^{\nu_2-1} x_3^{\nu_3-1}
 \frac{{\mathcal U}^{\nu-\frac{3}{2}D}}{{\mathcal F}^{\nu-D}} \omega
\end{align}
with the two Feynman graph polynomials
\begin{align}
 {\mathcal U} \; = \; x_1 x_2 + x_2 x_3 + x_3 x_1,
 & &
 {\mathcal F} \; = \; - x_1 x_2 x_3 t + m^2 \left( x_1 + x_2 + x_3 \right) {\mathcal U}.
\end{align}
The differential two-form $\omega$ is given by
\begin{align}
 \omega = x_1 dx_2 \wedge dx_3 + x_2 dx_3 \wedge dx_1 + x_3 dx_1 \wedge dx_2,
\end{align}
and the integration is over
\begin{align}
\label{def_sigma}
 \sigma = \left\{ \left[ x_1 : x_2 : x_3 \right] \in {\mathbb P}^2 | x_i \ge 0, i=1,2,3 \right\}.
\end{align}
Integration-by-parts identities \cite{Tkachov:1981wb,Chetyrkin:1981qh}
allow us to express any member of the Feynman integral family in terms 
of a few master integrals.
For the sunrise family we need three master integrals.
A possible basis of master integrals is given by
\begin{align}
\label{naive_basis_master_integrals}
 S_{110}(D,t), \; S_{111}(D,t), \; S_{211}(D,t).
\end{align}
The Feynman graphs for these integrals are shown in fig.~(\ref{fig_master_integrals}).
In the top-topology with three different propagators we have two master integrals ($S_{111}$ and $S_{211}$).
The integral $S_{110}$ has only two different propagators and corresponds to a subtopology.

Dimensional-shift relations \cite{Tarasov:1996br,Tarasov:1997kx}
relate integrals in $(D-2)$ space-time dimensions 
to integrals in $D$ space-time dimensions.
We may therefore work without loss of generality in $D=2-2\eps$ space-time dimensions.
The physical relevant expressions for $D=4-2\eps$ space-time dimensions may be obtained
from the ones for $2-2\eps$ space-time dimensions with the help of the dimensional-shift relations.
Let us pause for a moment and consider the sunrise integral $S_{111}$ around two space-time dimensions.
Working in $2-2\eps$ space-time dimensions is advantageous for two reasons:
First of all, the integral $S_{111}$ is finite for $D=2$.
Secondly -- and this is the main reason for working in $D=2-2\eps$ dimensions --
the lowest-order term of the $\eps$-expansion of $S_{111}(2-2\eps,t)$ depends only on the
graph polynomial ${\mathcal F}$, but not on ${\mathcal U}$.
Indeed, we have from eq.~(\ref{def_Feynman_integral})
\begin{align}
 S_{111}\left( 2, t\right)
 =
 \mu^2
 \int\limits_{\sigma} 
 \frac{\omega}{{\mathcal F}}.
\end{align}
The integral $S_{111}(2,t)$ depends on two geometric objects: The integration region $\sigma$
and the variety defined by the zero set of the polynomial ${\mathcal F}$:
\begin{align}
\label{def_zero_set_F}
 {\mathcal F} = 0.
\end{align}
Eq.~(\ref{def_zero_set_F}) is a polynomial of degree $3$ in the Feynman parameters $x_1$, $x_2$ and $x_3$
and defines together with the choice of a rational point an elliptic curve.
This elliptic curve will be discussed in more detail in section~\ref{sec:EllipticCurve}.

In order to present the system of differential equations for the set of master integrals it will be convenient
to rescale them and to take as a basis of master integrals the set
\begin{align}
\label{def_basis}
 I_1\left(t\right)
 & =
 \eps^2 S_{110}\left(2-2\eps,t\right),
 \nonumber \\
 I_2\left(t\right)
 & = 
 \frac{m^2}{\mu^2} S_{111}\left(2-2\eps,t\right),
 \nonumber \\
 I_3\left(t\right)
 & =
 \frac{m^4}{\mu^4} S_{211}\left(2-2\eps,t\right).
\end{align}
We denote the vector of basis integrals by $\vec{I}=(I_1,I_2,I_3)^T$.
Let us now consider the derivatives of the basis integrals with respect to the variable $t$.
These derivatives can be expressed again as a linear combination of the basis integrals.
This gives us a differential equation for the 
basis integrals \cite{Kotikov:1990kg,Kotikov:1991pm,Remiddi:1997ny,Gehrmann:1999as,Argeri:2007up,MullerStach:2012mp,Henn:2013pwa,Henn:2014qga,Adams:2017tga}:
The differential equation is of Fuchsian type, where the only singularities are at $t\in\{0,m^2,9m^2,\infty\}$.
We have
\begin{align}
\label{dgl_t}
 \mu^2 \frac{d}{dt} \vec{I}
 =
 \left[ 
  \frac{\mu^2}{t} A_0 + \frac{\mu^2}{t-m^2} A_1 + \frac{\mu^2}{t-9m^2} A_9 \right] \vec{I},
\end{align}
where the $3 \times 3$-matrices $A_0$, $A_1$ and $A_9$ are polynomials in $\eps$ with rational coefficients.
The matrices are given by
\begin{align}
 A_0
 & =
 \left(
 \begin{array}{lll}
 0 & 0 & 0 \\
 0 & -1 -2 \eps & 3 \\
 0 & -\frac{1}{3}-\frac{5}{3}\eps - 2 \eps^2 & 1+3\eps \\
 \end{array}
 \right),
 &
 A_1
 & = 
 \left(
 \begin{array}{lll}
 0 & 0 & 0 \\
 0 & 0 & 0 \\
 \frac{1}{4} & \frac{1}{4}+\frac{5}{4}\eps + \frac{3}{2} \eps^2 & -1-2\eps \\
 \end{array}
 \right),
 \nonumber \\
 A_9
 & =
 \left(
 \begin{array}{lll}
 0 & 0 & 0 \\
 0 & 0 & 0 \\
 -\frac{1}{4} & \frac{1}{12}+\frac{5}{12} \eps + \frac{1}{2} \eps^2 & -1-2\eps \\
 \end{array}
 \right).
\end{align}
The tadpole integral $I_1$ is $t$-independent and given to all orders in $\eps$ by
\begin{align}
\label{tadpole}
 I_1(t) 
 =
 I_1(0)
 =
 \Gamma\left(1+\eps\right)^2 \left( \frac{m^2}{\mu^2} \right)^{-2 \eps}.
\end{align}
It is easily checked that the $\eps$-expansion of the basis integrals $\vec{I}=(I_1,I_2,I_3)$ starts 
at $\eps^0$.
For the basis integrals $\vec{I}=(I_1,I_2,I_3)$ we write
\begin{align}
 I_k\left(t\right)
 =
 e^{-2\gamma_E \eps} \sum\limits_{j=0}^\infty \eps^j  I_k^{(j)}\left(t\right),
 \qquad
 k = 1, ..., 3,
\end{align}
where $\gamma_E$ is Euler's constant.
For the tadpole integral $I_1(t)$ we have with $L=\ln(m^2/\mu^2)$
\begin{align}
 I_1(t)
 =
 e^{-2\gamma_E \eps} \left\{
 1 - 2 L \eps + \left[ \zeta_2 + 2 L^2 \right] \eps^2
 - \left[ \frac{2}{3} \zeta_3 - 2 L \zeta_2 - \frac{4}{3} L^3 \right] \eps^3
 \right\}
 + {\mathcal O}\left(\eps^4\right).
\end{align}
The system of first-order differential equations has always a block-triangular structure induced by the subtopologies.
We may therefore first solve all master integrals for all subtopologies and only then solve the top topology.
For the sunrise integral there is only one subtopology, whose solution has been given in eq~(\ref{tadpole}).
Let us now turn to the top topology consisting of two master integrals $(I_2,I_3)$.
It will be convenient to re-write the two coupled first-order differential equations for the top-sector $(I_2,I_3)$ 
as an inhomogeneous second-order differential equation for a single master integral.
We may take $S_{111}(2-2\eps,t)$ or $I_2(t)$ as this master integral, but it is advantageous 
to use \footnote{Note that the prefactor differs slightly from our previous publication \cite{Adams:2015ydq}.
Here we use $\sqrt{-t}$, whereas in \cite{Adams:2015ydq} we used $\sqrt{t}$.
In this publication the base of the $\eps$-power is positive for $t<0$, whereas in \cite{Adams:2015ydq} it is 
positive for $0<t<m^2$.
The conventions in this publication are adapted to the Euclidean region $t<0$.
Since the final result is analytic in a neighbourhood of $t=0$, both approaches lead in a neighbourhood of $t=0$
to the same result.}
\begin{align}
\label{def_tilde_I2}
 \tilde{S}_{111}(2-2\eps,t)
 =
  \left[ \Gamma\left(1+\eps\right) \right]^{-2} 
 \left( \frac{3 \mu^4 \sqrt{-t}}{m \left(t-m^2\right) \left(t-9m^2\right)} \right)^{-\eps}
 S_{111}(2-2\eps,t).
\end{align}
The differential equation for the integral $\tilde{S}_{111}(2-2\eps,t)$ reads
\begin{align}
\label{diff_eq_I2tilde}
 L_2 \tilde{S}_{111}(2-2\eps,t)
 =
 - \frac{6 \mu^2}{t\left(t-m^2\right)\left(t-9m^2\right)} 
 \left( \frac{\left(t-m^2\right) \left(t-9m^2\right)}{3 m^3 \sqrt{-t}} \right)^{\eps},
\end{align}
where the Picard-Fuchs operator $L_2$ is given by
\begin{align}
 L_2 = L_2^{(0)} + \eps^2 L_2^{(2)},
\end{align}
with
\begin{align}
\label{coeff_picard_fuchs}
 L_2^{(0)} & =
  \frac{d^2}{dt^2} 
  + \left( \frac{1}{t} + \frac{1}{t-m^2} + \frac{1}{t-9m^2} \right) \frac{d}{dt} 
        + \frac{1}{m^2} \left( - \frac{1}{3 t} + \frac{1}{4\left(t-m^2\right)} + \frac{1}{12\left(t-9m^2\right)} \right),
 \nonumber \\
 L_2^{(2)} & =
            -\frac{\left(t+3m^2\right)^4}{4 t^2 \left(t-m^2\right)^2\left(t-9m^2\right)^2}.
\end{align}
The choice of the master integral in eq.~(\ref{def_tilde_I2}) ensures that the Picard-Fuchs operator
in eq.~(\ref{coeff_picard_fuchs}) has a particular nice form, i.e. the only terms involving higher powers of $\eps$
are the ones given by $\eps^2 L_2^{(2)}$. These terms do not involve any derivatives $d/dt$.

\section{The elliptic curve}
\label{sec:EllipticCurve}

We may view the graph polynomial ${\mathcal F}$ as a polynomial in the Feynman parameters 
$x_1$, $x_2$, $x_3$ with parameters $t$ and $m^2$.
The algebraic equation
\begin{align}
 {\mathcal F} = 0
\end{align}
defines together with the choice of a rational point as origin an elliptic curve.
Rational points are for example the three intersection points of the integration region $\sigma$ (defined in eq.~(\ref{def_sigma}))
with the variety defined by ${\mathcal F}=0$. These points are given by
\begin{align}
\label{intersection_F_sigma}
 P_1 = \left[1:0:0\right], 
 \;\;\;
 P_2 = \left[0:1:0\right], 
 \;\;\;
 P_3 = \left[0:0:1\right].
\end{align}
We may choose one of these three points $P_1$, $P_2$, $P_3$ as the origin $O$.
Let us take $O=P_3$.
The elliptic curve can be transformed into the Weierstrass normal form
\begin{align}
\label{WNF_with_g2_g3}
 y^2 z = 4 x^3 - g_2 x z^2 - g_3 z^3
\end{align}
by a birational transformation.
Explicitly, this transformation is given for $[x_1:x_2:x_3] \neq [0:0:1]$ by \cite{vanHoeij:1995}
\begin{align}
\label{weierstrass_trafo}
 x & = \frac{1}{\mu^4} \left[ \left(t-3m^2\right)^2 \left(x_1+x_2\right)^2 - 12 m^2 \left(t-m^2\right) \left(x_1+x_2\right) x_3 \right],
 \nonumber \\
 y & = - \frac{12 m^2 \left(t-m^2\right)}{\mu^6} \left(x_1-x_2\right) \left[ m^2 \left(x_1+x_2\right) - \left(t-m^2\right) x_3 \right],
 \nonumber \\
 z & = 12 \left(x_1+x_2\right)^2.
\end{align}
The inverse transformation is given for $[x:y:z] \neq [0:1:0]$ by
\begin{align}
 x_1 & = - \frac{6m^2}{\mu^8} z \left[ \left(t-m^2\right) \left(t^2-6m^2t - 3m^4\right) z - 12 \left(t-m^2\right) \mu^4 x + 12 \mu^6 y \right],
 \nonumber \\
 x_2 & = - \frac{6m^2}{\mu^8} z \left[ \left(t-m^2\right) \left(t^2-6m^2t - 3m^4\right) z - 12 \left(t-m^2\right) \mu^4 x - 12 \mu^6 y \right],
 \nonumber \\
 x_3 & = - \frac{1}{\mu^8} \left[ \left(t-3m^2\right)^2 z - 12 \mu^4 x \right]
                          \left[ \left(t^2-6m^2 t - 3 m^4 \right) z - 12 \mu^4 x \right].
\end{align}
The point $[x_1:x_2:x_3] = [0:0:1]$ is transformed to the point $[x:y:z]=[0:1:0]$.
Let us denote by $Q_1$ and $Q_2$ the images of the points $P_1$ and $P_2$, respectively.
$Q_1$ and $Q_2$ are given by
\begin{align}
 Q_1 = \left[ \frac{\left(t-3m^2\right)^2}{12 \mu^4} : - \frac{m^4\left(t-m^2\right)}{\mu^6} : 1 \right],
 \qquad
 Q_2 = \left[ \frac{\left(t-3m^2\right)^2}{12 \mu^4} : \frac{m^4\left(t-m^2\right)}{\mu^6} : 1 \right].
\end{align}
In the following we will work in the chart $z=1$.
Factorising the cubic polynomial on the right-hand side of eq.~(\ref{WNF_with_g2_g3}),
the Weierstrass normal form can equally be written as
\begin{align}
 y^2 = 4 \left(x-e_1\right)\left(x-e_2\right)\left(x-e_3\right),
 \;\;\;\;\;\;
 \mbox{with} 
 \;\;\;
 e_1+e_2+e_3=0,
\end{align}
and
\begin{align}
 g_2 = -4 \left( e_1 e_2 + e_2 e_3 + e_3 e_1 \right),
 \qquad
 g_3 = 4 e_1 e_2 e_3.
\end{align}
The roots are given by
\begin{align}
\label{def_roots}
 e_1 
 & = 
 \frac{1}{24 \mu^4} \left( -t^2 + 6 m^2 t + 3 m^4 + 3 \sqrt{\tilde{D}} \right),
 \nonumber \\
 e_2 
 & = 
 \frac{1}{24 \mu^4} \left( -t^2 + 6 m^2 t + 3 m^4 - 3 \sqrt{\tilde{D}} \right),
 \nonumber \\
 e_3 
 & = 
 \frac{1}{24 \mu^4} \left( 2 t^2 - 12 m^2 t - 6 m^4 \right).
\end{align}
As abbreviation we used
\begin{align}
\label{def_D}
 \tilde{D} & =
 \left( t - m^2 \right)^3
 \left( t - 9 m^2 \right).
\end{align}
The modulus $k$ and the complementary modulus $k'$ of the elliptic curve are defined by
\begin{align}
\label{def_modulus}
 k = \sqrt{\frac{e_3-e_2}{e_1-e_2}},
 \qquad
 k' = \sqrt{1-k^2} = \sqrt{\frac{e_1-e_3}{e_1-e_2}}.
\end{align}
The periods of the elliptic curve are given by
\begin{align}
\label{def_periods}
 &
\psi_1 =  
 2 \int\limits_{e_2}^{e_3} \frac{dx}{y}
 =
 \frac{4 \mu^2}{\tilde{D}^{\frac{1}{4}}} K\left(k\right),
 & &
 \psi_2 =  
 2 \int\limits_{e_1}^{e_3} \frac{dx}{y}
 =
 \frac{4 i \mu^2}{\tilde{D}^{\frac{1}{4}}} K\left(k'\right),
 \\
 &
 \phi_1 =  
 \frac{8\mu^4}{\tilde{D}^{\frac{1}{2}}} \int\limits_{e_2}^{e_3} \frac{\left(x-e_2\right) dx}{y}
 =
 \frac{4 \mu^2}{\tilde{D}^{\frac{1}{4}}} \left( K\left(k\right)- E\left(k\right) \right),
 & &
 \phi_2 =  
 \frac{8\mu^4}{\tilde{D}^{\frac{1}{2}}} \int\limits_{e_1}^{e_3} \frac{\left(x-e_2\right) dx}{y}
 =
 \frac{4 i \mu^2}{\tilde{D}^{\frac{1}{4}}} E\left(k'\right).
 \nonumber
\end{align}
$K(x)$ and $E(x)$ denote the complete elliptic integral of the first kind and second kind, respectively:
\begin{align}
 K(x)
 = 
 \int\limits_0^1 \frac{dt}{\sqrt{\left(1-t^2\right)\left(1-x^2t^2\right)}},
 \qquad
 E(x)
 = 
 \int\limits_0^1 dt \sqrt{\frac{1-x^2t^2}{1-t^2}}.
\end{align}
The derivatives of the periods with respect to the variable $t$ are given by
\begin{align}
\label{derivative_psi_phi}
 \frac{d}{dt} \left( \begin{array}{c} \psi_i \\ \phi_i \\ \end{array} \right)
 = B \left( \begin{array}{c} \psi_i \\ \phi_i \\ \end{array} \right),
 \qquad
 B = \frac{d}{dt} 
 \left( \begin{array}{cc}
 - \frac{1}{2} \ln Z_2 & \frac{1}{2} \ln \frac{Z_2}{Z_1} \\
 - \frac{1}{2} \ln \frac{Z_2}{Z_3} & \frac{1}{2} \ln \frac{Z_2}{Z_3^2}
 \end{array} \right),
\end{align}
with $i \in \{1,2\}$ and 
\begin{align}
 Z_1 = e_3-e_2, 
 \qquad
 Z_2 = e_1-e_3,
 \qquad
 Z_3 = e_1-e_2.
\end{align}
Eq.~(\ref{derivative_psi_phi}) gives a coupled system of two first-order differential equations.
Alternatively, we may give a a single second-order differential equation for $\psi_i$.
One finds
\begin{align}
 L_2^{(0)} \psi_i = 0,
\end{align}
where the second-order differential operator is given by eq.~(\ref{coeff_picard_fuchs}).
Thus, the two periods $\psi_1$ and $\psi_2$ are the homogeneous solutions of the $\eps^0$-part of the
differential operator for the integral $\tilde{S}_{111}(2-2\eps,t)$.
The Wronskian is given by
\bq
\label{def_Wronski}
 W & = &
 \psi_1 \frac{d}{dt} \psi_2 - \psi_2 \frac{d}{dt} \psi_1
 =
 -
 \frac{12 \pi i \mu^4}{t\left( t - m^2 \right)\left( t - 9 m^2 \right)}.
\eq
We denote the ratio of the two periods $\psi_2$ and $\psi_1$ by
\begin{align}
\label{def_tau}
 \tau 
 =
 \frac{\psi_2}{\psi_1}
\end{align}
and the nome by
\begin{align}
\label{def_nome}
 q_2 = e^{i\pi \tau}.
\end{align}
We continue to use $q=\exp(2\pi i \tau)$ for the square of the nome.
This notation is consistent with the notation used in the section on modular forms. 
We denote by $r_3$ the third root of unity
\begin{align}
 r_3 \;\; = \;\; e^{\frac{2 \pi i}{3}} 
 \;\; = \;\;
 \frac{1+i\sqrt{3}}{1-i\sqrt{3}}
 \;\; = \;\;
 - \frac{1}{2} + \frac{i}{2} \sqrt{3}.
\end{align}
$r_3$ and $r_3^{-1}$ are the images of $Q_1$ and $Q_2$ in the 
Jacobi uniformization ${\mathbb C}^\ast/{q^{{\mathbb Z}}}$
of the elliptic curve \cite{Adams:2014vja}.

\section{Integration kernels}
\label{sec:integration_kernels}

In this section we solve the differential equation~(\ref{diff_eq_I2tilde})
and express the integration kernels as modular forms.

The integral $\tilde{S}_{111}(2-2\eps,t)$ has a Taylor expansion in $\eps$,
which we write as
\begin{align}
\label{expansion_tilde_2D}
 \tilde{S}_{111}(2-2\eps,t)
 =
 \sum\limits_{j=0}^\infty \eps^j \tilde{S}_{111}^{(j)}(2,t).
\end{align}
We insert this expansion into the differential equation~(\ref{diff_eq_I2tilde}) 
and consider the coefficient of $\eps^j$. This gives us
a differential equation for $\tilde{S}_{111}^{(j)}(2,t)$. This differential equation will involve lower-order terms
$\tilde{S}_{111}^{(i)}(2,t)$ with $i<j$, but not higher-order terms $\tilde{S}_{111}^{(k)}(2,t)$ with $k>j$.
We may therefore successively solve these differential equations, starting with $\tilde{S}_{111}^{(0)}(2,t)$.
The differential equation for $\tilde{S}_{111}^{(j)}(2,t)$ reads
\begin{align}
\label{dgl_transformed}
 L^{(0)}_{2} \tilde{S}_{111}^{(j)}(2,t)
  = 
 \mu^2 I^{(j)}(t),
 \qquad 
 I^{(j)}(t) = I_a^{(j)}(t) + I_b^{(j)}(t),
\end{align}
where the inhomogeneous terms are given by
\begin{align}
 I_a^{(j)}(t) & =
 - \frac{6}{t \left(t-m^2\right)\left(t-9m^2\right)} \frac{1}{j!} \ln^j\left( \frac{\left(t-m^2\right)\left(t-9m^2\right)}{3 m^3 \sqrt{-t}} \right),
 \nonumber \\
 I_b^{(j)}(t) & =
 \frac{\left(t+3m^2\right)^4}{4 \mu^2 t^2 \left(t-m^2\right)^2\left(t-9m^2\right)^2} \tilde{S}_{111}^{(j-2)}(2,t).
\end{align}
We already know that the space of homogeneous solutions of the differential equation~(\ref{dgl_transformed})
is spanned by the two periods $\psi_1$ and $\psi_2$, defined in eq.~(\ref{def_periods}).
Note that we have $i \pi \psi_2 = \psi_1 \ln(q_2)$.
Thus, we may write the general solution for the inhomogeneous equation as
\begin{align}
\label{inhomogeneous_solution}
 \tilde{S}_{111}^{(j)}(2,t)
 & =
 C_1^{(j)}\left(t_0\right) \psi_1\left(t\right) + C_2^{(j)}\left(t_0\right) \psi_1\left(t\right) \ln\left(q_2(t)\right)
 + \tilde{S}_{\mathrm{special}}^{(j)}\left(t,t_0\right),
\end{align}
where $\tilde{S}_{\mathrm{special}}^{(j)}\left(t,t_0\right)$ is a special solution of the inhomogeneous differential
equation.
Variation of the constants gives us an expression for the special solution
\begin{align}
 \tilde{S}_{\mathrm{special}}^{(j)}\left(t,t_0\right)
 & = 
  \mu^2 \int\limits_{t_0}^{t} dt_1 \frac{I^{(j)}(t_1)}{W(t_1)} \left[ - \psi_1(t) \psi_2(t_1) + \psi_2(t) \psi_1(t_1) \right].
\end{align}
We then change the integration variable from $t=p^2$ to the nome $q_2=\exp(\pi i \tau)$, where $\tau$ is defined by eq.~(\ref{def_tau}).
In a neighbourhood of $t=0$ we may invert the relation and obtain \cite{Adams:2014vja}
\begin{align}
\label{modular_function_t}
 t & = 
 - 9 m^2 
 \frac{\eta\left(\tau\right)^4 \eta\left(\frac{3\tau}{2}\right)^4 \eta\left(6\tau\right)^4}
      {\eta\left(\frac{\tau}{2}\right)^4 \eta\left(2\tau\right)^4 \eta\left(3\tau\right)^4}.
\end{align}
Eq.~(\ref{modular_function_t}) is obtained as follows:
We consider the modular lambda function \cite{Chandrasekharan} defined by
\bq
 \lambda & = & k^2 \; = \; \frac{e_3-e_2}{e_1-e_2}.
\eq
We may either view $\lambda$ as a function of $t$
\bq
 \lambda & = & \frac{t^2-6m^2 t-3m^4 + \left(m^2-t\right)^{\frac{3}{2}} \left(9m^2-t\right)^{\frac{1}{2}}}
                    {2\left(m^2-t\right)^{\frac{3}{2}} \left(9m^2-t\right)^{\frac{1}{2}}},
\eq
or as a function of $\tau$ (or $q_2$)
\bq
  \lambda
 & = &
 16 \frac{\eta\left(\frac{\tau}{2}\right)^8 \eta\left(2\tau\right)^{16}}{\eta\left(\tau\right)^{24}}.
\eq
The point $t=0$ corresponds to $\tau=i\infty$ and $q_2=0$.
We expand $\lambda$ on the one hand as a function of $t$ around $t=0$, and on the other hand as a function of $q_2$
around $q_2=0$. This gives
\bq
 - \frac{16}{9} \frac{t}{m^2} - \frac{64}{27} \left(\frac{t}{m^2}\right)^2 - \frac{2080}{729} \left(\frac{t}{m^2}\right)^3 + ...
 & = &  
 16 q_2 - 128 q_2^2 + 704 q_2^3 + ...
\eq
Both Taylor series start at order $1$.
We may therefore use reversion on power series and obtain $t$ as a power series of $q_2$:
\bq
 t & = & - 9 m^2 \left( q_2 +4 q_2^2 + 10 q_2^3 + 20 q_2^4 + 39 q_2^5 + 76 q_2^6 + 140 q_2^7 + ... \right)
\eq
This can be done to high powers in $q_2$. We may then use the algorithms of \cite{Garvan:1998} to see if a representation in the form
of an eta quotient exists.
For the case at hand, the first $13$ coefficients are actually enough to find the eta quotient of eq.~(\ref{modular_function_t}).
We have checked that expanding up to ${\mathcal O}(q_2^{1000})$ will not alter our findings.

For the differential we have
\begin{align}
 dt & = \frac{\psi_1^2}{i \pi W} \frac{dq_2}{q_2}.
\end{align}
Partial integration leads to
\begin{align}
\label{special_solution}
 \tilde{S}_{\mathrm{special}}^{(j)}
 & = 
 -
 \frac{\psi_1}{\pi}
 \int\limits_{q_{2,0}}^{q_2} \frac{dq_2'}{q_2'}
 \int\limits_{q_{2,0}}^{q_2'} \frac{dq_2''}{q_2''}
 \;
 \frac{\psi_1(q_2'')^3}{\pi W(q_2'')^2} 
 \; 
 \mu^2 I^{(j)}(q_2'').
\end{align}
We obtain a two-fold integration over the inhomogeneous term multiplied by $\psi_1^3/(\pi W^2)$.
The function $\tilde{S}_{111}^{(j-2)}\left(2,t\right)$ contains always a prefactor $\psi_1/\pi$
and it is convenient to write
\begin{align}
\label{def_Etilde}
 \tilde{S}_{111}^{(j)}\left(2,t\right)
 =
 \frac{\psi_1}{\pi}
 \tilde{E}_{111}^{(j)}\left(2,q_2\right).
\end{align}
We would like to show that all functions $\tilde{E}_{111}^{(j)}\left(2,q\right)$ can be written as iterated
integrals of modular forms.
To this aim we investigate the integration kernels in more detail.
Let us start with the first term in the $\eps$-expansion. For $j=0$ only the term $I_a^{(0)}$ contributes.
Let us denote
\begin{align}
 f_3 & =
 \frac{\mu^2 \psi_1^3}{\pi W^2}
 \;
 \frac{6}{t \left(t-m^2\right)\left(t-9m^2\right)}.
\end{align}
We may express $f_3$ in terms of $\mathrm{ELi}$-functions, where the definition of the $\mathrm{ELi}$-functions
is given in appendix~\ref{sect:ELi}:
\begin{align}
 f_3 & =
 \frac{3}{i} \left[ \mathrm{ELi}_{0;-2}\left(r_3;-1;-q_2\right) - \mathrm{ELi}_{0;-2}\left(r_3^{-1};-1;-q_2\right) \right].
\end{align}
In addition, we have a representation in the form of an eta quotient:
\begin{align}
 f_3 & =
 3 \sqrt{3} 
 \frac{ \eta\left(\tau\right)^{11} \eta\left(3\tau\right)^{7} }
      { \eta\left(\frac{\tau}{2}\right)^{5} \eta\left(2\tau\right)^{5} \eta\left(\frac{3\tau}{2}\right) \eta\left(6\tau\right) }.
\end{align}
\ifthenelse{\boolean{arxiveversion}}
{
Setting $\tau_2=\tau/2$ we may now check with the help of the theorem of Ligozat 
from section~\ref{sect_eta_quotients} if $f_3$ corresponds to a modular form.
It does and we find
\begin{align}
 f_3(\tau_2) \in 
 \mathcal{M}_3(12,\chi_1).
\end{align}
We may therefore express $f_3$ as a linear combination of the basis elements of $\mathcal{M}_3(12,\chi_1)$.
We find with $q_2=\exp(2\pi i \tau_2) = \exp(i \pi \tau)$
\begin{align}
\label{def_f3}
 f_3(\tau_2)
 & =
 3 \sqrt{3} \left[
 E_3\left(\tau_2;\bar{\chi}_1,\bar{\chi}_0\right)
 + 2 E_3\left(2\tau_2;\bar{\chi}_1,\bar{\chi}_0\right)
 - 8 E_3\left(4\tau_2;\bar{\chi}_1,\bar{\chi}_0\right)
 \right].
\end{align}
} 
{
Let us introduce two primitive Dirichlet characters $\bar{\chi}_0$, $\bar{\chi}_1$,
defined in terms of the Kronecker symbol by
\begin{align}
\bar{\chi}_0 = \left( \dfrac{1}{n} \right), \qquad
\bar{\chi}_1 = \left( \dfrac{-3}{n} \right).
\end{align}
We denote by $\chi_0$ and $\chi_1$ the induced characters modulo $12$.
The conductors of the characters $\chi_0$ and $\chi_1$ are
$1$ and $3$, respectively.
Setting $\tau_2=\tau/2$ we may now check with the help of the theorem of Ligozat \cite{Ligozat:1975,Alaca:2015aa} 
if $f_3$ corresponds to a modular form.
It does and we find
\begin{align}
 f_3(\tau_2) \in 
 \mathcal{M}_3(12,\chi_1).
\end{align}
We may therefore express $f_3$ as a linear combination of the basis elements of $\mathcal{M}_3(12,\chi_1)$.
We find with $q_2=\exp(2\pi i \tau_2) = \exp(i \pi \tau)$
\begin{align}
\label{def_f3}
 f_3(\tau_2)
 & =
 3 \sqrt{3} \left[
 E_3\left(\tau_2;\bar{\chi}_1,\bar{\chi}_0\right)
 + 2 E_3\left(2\tau_2;\bar{\chi}_1,\bar{\chi}_0\right)
 - 8 E_3\left(4\tau_2;\bar{\chi}_1,\bar{\chi}_0\right)
 \right].
\end{align}
The notation for the generalised Eisenstein series is given in appendix~\ref{sect:eisenstein_space}.
} 
Let us now go to higher orders in $\eps$.
For $j>0$ we obtain from $I_a^{(j)}$ in addition to the modular form $f_3$ powers of the logarithm
\begin{align}
 L_{\mathrm{inhom}}
 & = 
 \ln\left( \frac{\left(t-m^2\right)\left(t-9m^2\right)}{3 m^3 \sqrt{-t}} \right).
\end{align}
We may re-write the logarithm as an integral over a modular form
\begin{align}
 L_{\mathrm{inhom}}
 & = 
 I\left(f_2;q_2\right),
\end{align}
where $f_2 \in \mathcal{M}_2(12,\chi_0)$ is given by
\ifthenelse{\boolean{arxiveversion}}
{
\begin{align}
\label{def_f2}
 f_2(\tau_2)
 & =
 14 B_{2,2}(\tau_2) - 4 B_{2,3}(\tau_2) - 8 B_{2,4}(\tau_2) + 10 B_{2,6}(\tau_2) - 4 B_{2,12}(\tau_2).
\end{align}
} 
{
\begin{align}
\label{def_f2}
 f_2(\tau_2)
 & =
 14 B_{2,2}(\tau_2) - 4 B_{2,3}(\tau_2) - 8 B_{2,4}(\tau_2) + 10 B_{2,6}(\tau_2) - 4 B_{2,12}(\tau_2).
\end{align}
The Eisenstein series $B_{2,K}(\tau_2)$ are defined in appendix~\ref{sect:eisenstein_space}.
} 
For the $j$-th power of the logarithm we may use the shuffle product to re-write this expression
as a single iterated integral:
\begin{align}
 \frac{1}{j!}
 \left( L_{\mathrm{inhom}} \right)^j
 & =
 I(\{f_2\}^j;q_2)
 =
 I(\underbrace{f_2,...,f_2}_{j};q_2).
\end{align}
Let us now look at the inhomogeneous parts coming from $I_b^{(j)}$.
We have to consider
\begin{align}
 f_4
 & =
 - \frac{\mu^2 \psi_1^3}{\pi W^2} 
 \;
 \frac{\left(t+3m^2\right)^4}{4 \mu^2 t^2 \left(t-m^2\right)^2\left(t-9m^2\right)^2} 
 \; \frac{\psi_1}{\pi},
\end{align}
where the additional factor of $\psi_1/\pi$ comes from eq.~(\ref{def_Etilde}).
We have
\begin{align}
 f_4 & = f_1^4,
\end{align}
with
\begin{align}
\label{def_f1}
 f_1 & = \frac{\left(t+3m^2\right)}{2 \sqrt{6} \mu^2} \; \frac{\psi_1}{\pi}.
\end{align}
$f_1$ is a modular form of weight $1$ and $f_1 \in \mathcal{M}_1(12,\chi_1)$. We have
\begin{align}
 f_1 & =
 -3 \sqrt{2} \left[ E_1(\tau_2,\bar{\chi}_0,\bar{\chi}_1) - 2 E_1(4 \tau_2,\bar{\chi}_0,\bar{\chi}_1) \right].
\end{align}
Thus $f_4 \in \mathcal{M}_4(12,\chi_0)$ since $\chi_1^2 = \chi_0$ and hence $\chi_1^4=\chi_0$.

From eqs.~(\ref{def_f3}), (\ref{def_f2}) and (\ref{def_f1})
we see that all integration kernels are modular forms with characters for $\Gamma_0(12)$.
It follows that $\tilde{E}_{111}^{(j)}$ can be written as a linear combination of iterated integrals of modular forms.

Let us make a small detour towards the kite integral.
The kite integral is a two-loop two-point function with five internal propagators, 
\begin{figure}
\begin{center}
\includegraphics[scale=1.0]{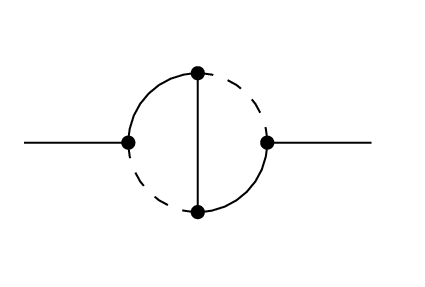}
\end{center}
\caption{
\it The kite integral. Solid lines correspond to massive propagators, dashed lines to massless propagators.
}
\label{fig_kite}
\end{figure}
shown in 
fig.~(\ref{fig_kite}).
Three internal propagators are massive with mass $m$, two internal propagators are massless.
The kite integral contains the sunset integral as subtopology.
The kite integral can be expressed to all orders in $\eps$ in terms of $\mathrm{ELi}$-functions,
an algorithm to obtain the $\eps^j$-term in the $\eps$-expansion has been given in \cite{Adams:2016xah}.
For the discussion of the kite integral we follow the notation of ref.~\cite{Adams:2016xah}.
We may now ask, if the kite integral can be expressed at each order in the $\eps$-expansion as a linear
combination of iterated integrals of modular forms.
This is indeed the case. In order to show this, all what needs to be done is to express the additional
integration kernels in terms of modular forms.
For the kite integral we have in addition to the integration kernels already present in the sunrise integral,
five additional integration kernels.
Three of them are modular forms of weight $2$.
We denote them as $g_{2,0}, g_{2,1}, g_{2,9} \in \mathcal{M}_2(12,\chi_0)$.
They are given by
\begin{align}
 g_{2,0}
 & =
 \frac{1}{i \pi} \frac{\psi_1^2}{W} \frac{1}{t}
 & = &
 4 \left[ B_{2,2}(\tau_2) + B_{2,3}(\tau_2) - B_{2,4}(\tau_2) - B_{2,6}(\tau_2) + B_{2,12}(\tau_2) \right], 
 \nonumber \\
 g_{2,1}
 & =
 \frac{1}{i \pi} \frac{\psi_1^2}{W} \frac{1}{t-m^2}
 & = &
 3 \left[ 6 B_{2,2}(\tau_2) + B_{2,3}(\tau_2) - 3 B_{2,4}(\tau_2) - 2 B_{2,6}(\tau_2) + B_{2,12}(\tau_2) \right], 
 \nonumber \\
 g_{2,9}
 & =
 \frac{1}{i \pi} \frac{\psi_1^2}{W} \frac{1}{t-9m^2}
 & = &
 - 2 B_{2,2}(\tau_2) - 5 B_{2,3}(\tau_2) - B_{2,4}(\tau_2) + 14 B_{2,6}(\tau_2) - 5 B_{2,12}(\tau_2).
\end{align}
The remaining two additional integration kernels are modular forms of weight $3$, which we denote as
$g_{3,0}, g_{3,1} \in \mathcal{M}_3(12,\chi_1)$.
They are given by
\begin{align}
 g_{3,0}
 & =
 \frac{1}{i \pi \mu^2} \frac{\psi_1^2}{W} \frac{\psi_1}{\pi}
 & = &
 - 2 f_3,
 \nonumber \\
 g_{3,1}
 & =
 \frac{1}{i \pi \mu^2} \frac{\psi_1^2}{W} \frac{\psi_1}{\pi} \frac{t}{t-m^2}
 & = &
 -54 \sqrt{3} E_3\left(2\tau_2;\bar{\chi}_1,\bar{\chi}_0\right).
 \hspace*{50mm}
\end{align}
Since all integration kernels are modular forms it follows that the kite integral
can be written as a linear combination of iterated integrals of modular forms.
We have collected useful formulae for all modular forms relevant to the sunrise integral and the kite integral
in appendix~\ref{sect:summary_modular_forms}.

Let us now return to the two-loop sunrise integral. 
From eq.~(\ref{inhomogeneous_solution}) and eq.~(\ref{special_solution})
we easily obtain $\tilde{E}_{111}^{(j)}(2,q_2)$, order by order in $\eps$.
We have
\begin{align}
\label{Etilde_integral_eq}
 \tilde{E}_{111}^{(j)}(2,q_2)
 & =
 C_1^{(j)} + C_2^{(j)} \ln(q_2)
 +
 \int\limits_{0}^{q_2} \frac{dq_2'}{q_2'}
 \int\limits_{0}^{q_2'} \frac{dq_2''}{q_2''}
 \; \left[
 f_3 I\left(\left\{f_2\right\}^j;q_2''\right)
 + f_4 \tilde{E}_{111}^{(j-2)}(2,q_2'')
 \right],
\end{align}
where $C_1^{(j)}$ and $C_2^{(j)}$ are integration constants chosen in such a way as to match the boundary value 
at $q_2=0$. 
The first few terms for $\tilde{E}_{111}^{(j)}$ are given by
\begin{align}
 \tilde{E}_{111}^{(0)}(2,q_2)
 = & \;
 C_1^{(0)} + C_2^{(0)} \ln(q_2) + I(1,f_3;q_2),
 \nonumber \\
 \tilde{E}_{111}^{(1)}(2,q_2)
 = & \;
 C_1^{(1)} + C_2^{(1)} \ln(q_2) + I(1,f_3,f_2;q_2),
 \nonumber \\
 \tilde{E}_{111}^{(2)}(2,q_2)
 = &\;
 C_1^{(2)} + C_2^{(2)} \ln(q_2) + C_1^{(0)} I(1,f_4;q_2) + C_2^{(0)} I(1,f_4,1;q_2) + I(1,f_3,f_2,f_2;q_2) 
 \nonumber \\
 & \;
 + I(1,f_4,1,f_3;q_2).
\end{align}
We may then put all pieces together and obtain the solution for the sunrise integral
$S_{111}(2-2\eps,t)$.
We recall that we defined the Taylor expansions by
\begin{align}
 S_{111}\left(2-2\eps,t\right)
 & =
 e^{-2\gamma_E \eps} \sum\limits_{j=0}^\infty \eps^j S_{111}^{(j)}\left(2,t\right),
 &
 \tilde{S}_{111}(2-2\eps,t)
 & =
 \sum\limits_{j=0}^\infty \eps^j \tilde{S}_{111}^{(j)}(2,t).
\end{align}
It will be convenient to factor out $\psi_1/\pi$, therefore we set
\begin{align}
\label{def_E111}
 S_{111}\left(2-2\eps,t\right)
 & =
 \frac{\psi_1}{\pi}
 E_{111}(2-2\eps,q_2)
 =
 \frac{\psi_1}{\pi}
 e^{-2\gamma_E \eps} \sum\limits_{j=0}^\infty \eps^j E_{111}^{(j)}\left(2,q_2\right),
 \nonumber \\
 \tilde{S}_{111}(2-2\eps,t)
 & =
 \frac{\psi_1}{\pi}
 \tilde{E}_{111}(2-2\eps,q_2)
 =
 \frac{\psi_1}{\pi}
 \sum\limits_{j=0}^\infty \eps^j \tilde{E}_{111}^{(j)}(2,q_2).
\end{align}
In the Taylor expansion of $S_{111}(2-2\eps,t)$ and $E_{111}(2-2\eps,t)$
we have factored out a prefactor $\exp(-2\gamma_E \eps)$. This ensures that in the Taylor coefficients
$S_{111}^{(j)}(2,t)$ and $E_{111}^{(j)}(2,t)$ Euler's constant $\gamma_E$ does not appear.
$S_{111}(2-2\eps,t)$ and $\tilde{S}_{111}(2-2\eps,t)$ 
(and $E_{111}(2-2\eps,t)$ and $\tilde{E}_{111}(2-2\eps,t)$) are related by
\begin{align}
 S_{111}\left(2-2\eps,t\right)
 & =
 \Gamma\left(1+\eps\right)^2 e^{-2 \eps L -\eps I(f_2;q_2)} \tilde{S}_{111}\left(2-2\eps,q_2\right),
 \nonumber \\
 E_{111}\left(2-2\eps,t\right)
 & =
 \Gamma\left(1+\eps\right)^2 e^{-2 \eps L -\eps I(f_2;q_2)} \tilde{E}_{111}\left(2-2\eps,q_2\right),
\end{align}
with $L=\ln(m^2/\mu^2)$.
The factor $(\Gamma(1+\eps))^2$ ensures that also the Taylor coefficients
$\tilde{S}_{111}^{(j)}(2,t)$ and $\tilde{E}_{111}^{(j)}(2,t)$ are free of Euler's constant $\gamma_E$.
Note that we have
\begin{align}
 e^{\gamma_E \eps} \Gamma\left(1+\eps\right)
 & =
 \exp\left(\sum\limits_{j=2}^\infty \frac{\left(-1\right)^j}{j} \zeta_j \eps^j \right).
\end{align}
In order to obtain the full solution we have to specify some boundary values.
It will be convenient to do this at $t=0$ (corresponding to $q_2=0$).
The homogeneous solution $\psi_1$ has at $t=0$ the value
\begin{align}
 \frac{\psi_1(0)}{\pi}
 & =
 \frac{2 \mu^2}{\sqrt{3} m^2}.
\end{align} 
The boundary values for the two-loop sunrise integral can be obtained 
to any order in $\eps$ from the expansion of \cite{Adams:2015ydq}
\begin{align}
 \sum\limits_{j=0}^\infty \eps^j S_{111}^{(j)}\left(2,0\right)
 & =
 e^{2 \gamma \eps} \Gamma\left(1+2\eps\right)
 \left( \frac{m^2\sqrt{3}}{\mu^2} \right)^{-1-2\eps}
 \left[ \frac{3}{2\eps^2} \frac{\Gamma\left(1+\eps\right)^2}{\Gamma\left(1+2\eps\right)}
        h
        - \frac{\pi}{\eps} \right],
\end{align}
where
\begin{align}
 h
 & = 
 \frac{1}{i}
 \left[ 
  \left(-r_3\right)^{-\eps} \; {}_2F_1\left(-2\eps,-\eps;1-\eps; r_3 \right)
  -
  \left(-r_3^{-1}\right)^{-\eps} \; {}_2F_1\left(-2\eps,-\eps;1-\eps; r_3^{-1} \right)
 \right].
\end{align}
The hypergeometric function can be expanded systematically in $\eps$
with the methods of \cite{Moch:2001zr}.
The first few terms are given by
\begin{align}
 {}_2F_1\left(-2\eps,-\eps;1-\eps; x \right)
 & =
 1 + 2 \eps^2 \mathrm{Li}_2\left(x\right)
 + \eps^3 \left[ 2 \mathrm{Li}_3\left(x\right) - 4 \mathrm{Li}_{2,1}\left(x,1\right) \right]
 \nonumber \\
 &
 + \eps^4 \left[ 2 \mathrm{Li}_4\left(x\right) - 4 \mathrm{Li}_{3,1}\left(x,1\right) + 8 \mathrm{Li}_{2,1,1}\left(x,1,1\right) \right]
 + {\mathcal O}\left(\eps^5\right).
\end{align}
The first few boundary values for the sunrise integral are given by
\begin{align}
 S_{111}^{(0)}\left(2,0\right)
 & = 
 \frac{\sqrt{3} \mu^2}{i m^2}
 \left[ \mathrm{Li}_2\left(r_3\right) - \mathrm{Li}_2\left(r_3^{-1}\right) \right],
 \nonumber \\
 S_{111}^{(1)}\left(2,0\right)
 & = 
 \frac{\sqrt{3} \mu^2}{i m^2}
 \left\{
  - 2 \mathrm{Li}_{2,1}\left(r_3,1\right) - \mathrm{Li}_3\left(r_3\right) 
  + 2  \mathrm{Li}_{2,1}\left(r_3^{-1},1\right) + \mathrm{Li}_3\left(r_3^{-1}\right)
 \right\}
 \nonumber \\
 & 
  - 2 \ln\left(\frac{m^2 \sqrt{3}}{\mu^2}\right) S_{111}^{(0)}\left(2,0\right),
 \nonumber \\
 S_{111}^{(2)}\left(2,0\right)
 & = 
 \frac{\sqrt{3} \mu^2}{i m^2}
 \left\{
    4 \mathrm{Li}_{2,1,1}\left(r_3,1,1\right) - 2 \mathrm{Li}_{3,1}\left(r_3,1\right) + \mathrm{Li}_4\left(r_3\right) 
  - 4 \mathrm{Li}_{2,1,1}\left(r_3^{-1},1,1\right) 
 \right. 
 \nonumber \\
 & 
 \left.
  + 2 \mathrm{Li}_{3,1}\left(r_3^{-1},1\right) 
  - \mathrm{Li}_4\left(r_3^{-1}\right) 
  + \frac{2\pi^2}{9} \left[
                           \mathrm{Li}_2\left(r_3\right) - \mathrm{Li}_2\left(r_3^{-1}\right) 
                    \right]
 \right\}
 \nonumber \\
 &  
  - 2 \ln\left(\frac{m^2 \sqrt{3}}{\mu^2}\right) 
        S_{111}^{(1)}\left(2,0\right)
  - 2 \ln^2\left(\frac{m^2 \sqrt{3}}{\mu^2}\right) S_{111}^{(0)}\left(2,0\right).
\end{align}
The multiple polylogarithms are defined by
\bq
\label{def_multiple_polylogs}
 \mathrm{Li}_{n_1,n_2,...,n_l}\left(x_1,x_2,...,x_l\right)
 & = &
 \sum\limits_{j_1=1}^\infty \sum\limits_{j_2=1}^{j_1-1} ... \sum\limits_{j_l=1}^{j_{l-1}-1}
 \frac{x_1^{j_1}}{j_1^{n_1}} \frac{x_2^{j_2}}{j_2^{n_2}} ... \frac{x_l^{j_l}}{j_l^{n_l}}.
\eq
In accordance with eq.~(\ref{def_E111}) we set
\begin{align}
 E_{111}^{(j)}(2,0) 
 & =
 \frac{\pi}{\psi_1(0)} 
 S_{111}^{(j)}(2,0)
 =
 \frac{\sqrt{3}{m^2}}{2 \mu^2} 
 S_{111}^{(j)}(2,0).
\end{align}
Putting everything together we obtain the solution of the two-loop sunrise integral.
The first few terms in the $\eps$-expansion are given by
\begin{align}
 S_{111}^{(0)}(2,t)
 & =
 \frac{\psi_1}{\pi}
 \left[
  E_{111}^{(0)}(2,0) + I\left(1,f_3;q_2\right)
 \right],
 \nonumber \\
 S_{111}^{(1)}(2,t)
 & =
 \frac{\psi_1}{\pi}
 \left[
  E_{111}^{(1)}(2,0) 
  - \left( I\left(f_2;q_2\right) + \frac{1}{2} I\left(1;q_2\right) \right) E_{111}^{(0)}(2,0) 
  + I\left(1,f_3,f_2;q_2\right)
 \right. \nonumber \\
 & \left.
  - I\left(f_2;q_2\right) I\left(1,f_3;q_2\right)
  - 2 L \; I\left(1,f_3;q_2\right)
 \right],
 \nonumber \\
 S_{111}^{(2)}(2,t)
 & =
 \frac{\psi_1}{\pi}
 \left[
  E_{111}^{(2)}(2,0) 
  - \left( I\left(f_2;q_2\right) + \frac{1}{2} I\left(1;q_2\right) \right) E_{111}^{(1)}(2,0) 
 \right. \nonumber \\
 & \left.
  + \left( I\left(1,f_4;q_2\right) + I\left(f_2,f_2;q_2\right) + \frac{1}{2} I\left(1;q_2\right) I\left(f_2;q_2\right) \right) E_{111}^{(0)}(2,0) 
 \right. \nonumber \\
 & \left.
 + I\left(1,f_3,f_2,f_2;q_2\right) + I\left(1,f_4,1,f_3;q_2\right) - I\left(f_2;q_2\right) I\left(1,f_3,f_2;q_2\right) 
 \right. \nonumber \\
 & \left.
 + \left( \zeta_2 + I\left(f_2,f_2;q_2\right) \right) I\left(1,f_3;q_2\right)
 \right. \nonumber \\
 & \left.
 - 2 L 
  \left( I\left(1,f_3,f_2;q_2\right) - I\left(f_2;q_2\right) I\left(1,f_3;q_2\right) \right)
 + 2 L^2 I\left(1,f_3;q_2\right)
 \right].
\end{align}
We observe that the individual terms $S_{111}^{(j)}(2,t)$ have uniform depth.

\section{Modular transformations of the first term in the $\eps$-expansion}
\label{sec:TrafoSunrise}

We have seen in the previous section that each term of the $\eps$-expansion 
of the sunrise integral and the kite integral can be expressed as an iterated integral of modular forms.
The properties under modular transformations can be deduced from this representation.
The transformation properties are particularly simple for the first term in the $\eps$-expansion
of the sunrise integral around two space-time dimensions.
This is due to the fact that this term is an Eichler integral.
Concretely, $I(1,f_3;q_2)$ is an Eichler integral as defined by eqs.~(\ref{def_F_Eichler})-(\ref{F_Eichler}).
In this section we discuss as an example the sunrise integral in two space-time dimensions, i.e. the first term in the 
$\eps$-expansion around two space-time dimensions.
We show explicitly, how the period polynomial can be obtained.
The modular transformation properties of the sunrise integral in two space-time dimensions
have also been discussed in \cite{Bloch:2013tra}.

The integral is given by
\begin{align}
 S_{111}^{(0)}(2,t)
 & =
 \frac{\psi_1(q_2)}{\pi} E_{111}^{(0)}(2,q_2),
\end{align}
with
\begin{align}
 E_{111}^{(0)}(2,q_2) = E_{111}^{(0)}(2,0) + I\left(1,f_3;q_2\right)
 = 3 \;
 \mathrm{E}_{2;0}\left(r_3;-1;-q_2\right).
\end{align} 
The $\mathrm{E}$-function are defined in appendix~\ref{sect:ELi}.
We may view this integral either as a function of $t$, $q_2$ or $\tau_2$.
In this section we consider the integral as a function of $\tau_2$ and investigate the transformation
properties under the group $\Gamma_0(12)$.
The group $\Gamma_0(12)$ is generated by the elements
\begin{align}
\left\{
 \left(\begin{array}{rr}
 1 & 1 \\
 0 & 1 \\
 \end{array} \right),
 \left(\begin{array}{rr}
 7 & -1 \\
 36 & -5 \\
 \end{array} \right),
 \left(\begin{array}{rr}
 19 & -4 \\
 24 & -5 \\
 \end{array} \right),
 \left(\begin{array}{rr}
 17 & -5 \\
 24 & -7 \\
 \end{array} \right),
 \left(\begin{array}{rr}
  7 & -3 \\
 12 & -5 \\
 \end{array} \right),
 \left(\begin{array}{rr}
 -1  & 0 \\
 0 & -1 \\
 \end{array} \right)
 \right\}.
\end{align}
With a slight abuse of notation let us denote
\begin{align}
 \psi_1(\tau_2) = \psi_1(q_2(\tau_2)),
 \quad
 E_{111}^{(0)}(2,\tau_2) = E_{111}^{(0)}(2,q_2(\tau_2)),
 \quad
 S_{111}^{(0)}(2,\tau_2) = \frac{\psi_1(\tau_2)}{\pi} E_{111}^{(0)}(2,\tau_2).
\end{align}
As we show in appendix~\ref{sect:summary_modular_forms}, the function $\psi_1(\tau_2)$ is a modular form of weight $1$ for $\Gamma_0(12)$
with character $\chi_1$: 
$\psi_1 \in \mathcal{M}_1(12,\chi_1)$.
Thus
\begin{align}
 \psi_1\left(\gamma(\tau_2)\right)
 = \chi_1(d) \left(c\tau_2 +d \right) \psi_1\left(\tau_2\right)
 \qquad \text{for all} \;\; \gamma \in \Gamma_0(12).
\end{align}
$E_{111}^{(0)}(2,\tau_2)$ is an Eichler integral of $f_3 \in \mathcal{M}_3(12,\chi_1)$
and transforms therefore as
\begin{align}
 E_{111}^{(0)}\left(2,\gamma(\tau_2)\right)
 = \chi_1(d) \left(c\tau_2 +d \right)^{-1} \left[ E_{111}^{(0)}\left(2,\tau_2\right) + P_\gamma\left(\tau_2\right) \right]
 \qquad \text{for all} \;\; \gamma \in \Gamma_0(12),
\end{align}
where $P_\gamma(\tau_2)$ is a linear polynomial in $\tau_2$:
\begin{align}
 P_\gamma & = a_\gamma \tau_2 + b_\gamma.
\end{align}
It follows that $S_{111}^{(0)}(2,\tau_2)$ transforms as (note that $\chi_1^2=\chi_0$)
\begin{align}
 S_{111}^{(0)}\left(2,\gamma\left(\tau_2\right)\right) 
 = 
 \chi_0(d) \left[ S_{111}^{(0)}\left(2,\tau_2\right) + \frac{\psi_1\left(\tau_2\right)}{\pi} P_\gamma\left(\tau_2\right) \right]
 \qquad \text{for all} \;\; \gamma \in \Gamma_0(12).
\end{align}
The period polynomial $P_\gamma$ for $\gamma \in \Gamma_0(12)$ can be obtained as follows:
One chooses two values $\tau_2^{(a)}$ and $\tau_2^{(b)}$
and computes $E_{111}^{(0)}(2,\tau_2^{(a)})$, $E_{111}^{(0)}(2,\gamma(\tau_2^{(a)}))$, $E_{111}^{(0)}(2,\tau_2^{(b)})$
and $E_{111}^{(0)}(2,\gamma(\tau_2^{(b)}))$.
The coefficients of the period polynomial are then obtained by solving the linear system
\begin{align}
 a_\gamma \tau_2^{(a)} + b_\gamma & = \chi_1(d)^{-1} \left( c \tau_2^{(a)} + d \right) E_{111}^{(0)}(2,\gamma(\tau_2^{(a)})) - E_{111}^{(0)}(2,\tau_2^{(a)}),
 \nonumber \\
 a_\gamma \tau_2^{(b)} + b_\gamma & = \chi_1(d)^{-1} \left( c \tau_2^{(b)} + d \right) E_{111}^{(0)}(2,\gamma(\tau_2^{(b)})) - E_{111}^{(0)}(2,\tau_2^{(b)}).
\end{align}
Let us look at an example. We consider
\begin{align}
 \gamma_a & =
 \left(\begin{array}{rr}
 7 & -1 \\
 36 & -5 \\
 \end{array} \right).
\end{align}
Choosing $\tau_2^{(a)}=1/6$ and $\tau_2^{(b)}=(5+i)/36$
one obtains for the period polynomial
\begin{align}
 P_{\gamma_a}\left(\tau_2\right)
 & =
 9
 \left[ 
 \left(1+i\right)
  \mathrm{E}_{2;0}\left(r_3;-1;-e^{\frac{\pi i (5+i)}{18}}\right)
 + \left(1-i\right)
  \mathrm{E}_{2;0}\left(r_3;-1;-e^{\frac{\pi i (7+i)}{18}}\right)
 \right]
 \left( 6 \tau_2 - 1 \right)
 \nonumber \\
 & =
 4 i \pi^2 
 \left( 6 \tau_2 - 1 \right)
\end{align}
and $S_{111}^{(0)}(2,\tau_2)$ transforms under $\gamma_a$ as
\begin{align}
 S_{111}^{(0)}(2,\gamma_a(\tau_2))
 & =
 S_{111}^{(0)}(2,\tau_2)
 + \frac{\psi_1\left(\tau_2\right)}{\pi} P_{\gamma_a}\left(\tau_2\right).
\end{align}

\section{A compact expression to all orders}
\label{sec:all_order}

In section~\ref{sec:integration_kernels} we showed that the solution for the two-loop sunrise can be expressed 
as a linear combination of iterated integrals of modular forms. 
Each term in the $\eps$-expansion of the sunrise integral is obtained from simple integrations within this class of functions.
In this section we show that we may even give a generating function for the $j$-th term in the $\eps$-expansion.
Let us consider in more detail the function $\tilde{E}_{111}(2-2\eps,q_2)$, defined in eq.~(\ref{def_E111}).
From eq.~(\ref{Etilde_integral_eq}) it follows that $\tilde{E}_{111}(2-2\eps,q_2)$ satisfies the differential equation
\begin{align}
\label{diff_eq_Etilde}
 \left[
  \left( q_2 \frac{d}{dq_2} \right)^2 - \eps^2 f_4 \right] \tilde{E}_{111}\left(2-2\eps,q_2\right)
 & = 
 f_3 e^{\eps I(f_2;q_2)}.
\end{align}
Previously we solved this differential equation order by order in $\eps$.
This allows us to treat the expression $\eps^2 f_4 \tilde{E}_{111}(2-2\eps,q_2)$
as part of the inhomogeneous term.
The differential operator is then just
\begin{align}
 \left( q_2 \frac{d}{dq_2} \right)^2,
\end{align}
whose homogeneous solutions are simply
\begin{align}
 1, \qquad \ln(q_2).
\end{align}
In order to find the generating function for the $j$-th term of the $\eps$-expansion we now consider the differential operator
\begin{align}
 \left( q_2 \frac{d}{dq_2} \right)^2 - \eps^2 f_4.
\end{align}
The homogeneous solutions for this differential operator are
\begin{align}
 H_1 =
 \sum\limits_{j=0}^\infty \eps^{2j} I\left(\left\{1,f_4\right\}^j;q_2 \right),
 \qquad
 H_2 =
 \sum\limits_{j=0}^\infty \eps^{2j} I\left(\left\{1,f_4\right\}^j,1;q_2 \right).
\end{align}
The Wronskian equals
\begin{align}
 H_1 \frac{d H_2}{d \ln q_2} - H_2 \frac{d H_1}{d \ln q_2} 
 & = 1.
\end{align}
It is easily verified that a special solution for the differential equation~(\ref{diff_eq_Etilde})
is given by
\begin{align}
  \sum\limits_{j=0}^\infty 
   \eps^j 
   \sum\limits_{k=0}^{\lfloor \frac{j}{2} \rfloor} I\left( \left\{1,f_4\right\}^k, 1, f_3, \left\{f_2\right\}^{j-2k}; q_2\right),
\end{align}
where $\lfloor a \rfloor$ denotes the largest integer smaller or equal to $a$.
This allows us to write down a generating function for the $j$-th term of the $\eps$-expansion of the sunrise integral.
In oder to present this formula let us first define boundary values $B^{(j)}(2,0)$ through
\begin{align}
\label{def_boundary_values}
 S_{111}(2-2\eps,0) &= 
 \frac{\psi_1(0)}{\pi} \Gamma(1+\eps)^2 e^{-2\eps L} 
 \sum\limits_{j=0}^\infty \eps^j B^{(j)}(2,0),
\end{align}
i.e. we factor off from $S_{111}(2-2\eps,0)$ the normalised period $\psi_1(0)/\pi$ and the terms
\begin{align}
 \Gamma(1+\eps)^2 e^{-2\eps L} 
 & =
 e^{-2\eps L -2\gamma_E \eps + 2\sum\limits_{n=2}^\infty \frac{\left(-1\right)^n}{n} \zeta_n \eps^n}.
\end{align}
We obtain
\begin{align}
\label{main_result}
 S_{111}\left(2-2\eps,t\right)
 &= 
 \frac{\psi_1(q_2)}{\pi}
 e^{-\eps I(f_2;q_2) -2\eps L -2\gamma_E \eps + 2\sum\limits_{n=2}^\infty \frac{\left(-1\right)^n}{n} \zeta_n \eps^n} 
 \nonumber \\
 &
 \left\{
 \left[
 \sum\limits_{j=0}^\infty 
 \left(
 \eps^{2j} I\left(\left\{1,f_4\right\}^j;q_2 \right)
 -
 \frac{1}{2} \eps^{2j+1} I\left(\left\{1,f_4\right\}^j,1;q_2 \right)
 \right)
 \right]
 \sum\limits_{k=0}^\infty \eps^k B^{(k)}\left(2,0\right)
 \right. \nonumber \\
 & 
 \left.
 +
  \sum\limits_{j=0}^\infty 
   \eps^j 
   \sum\limits_{k=0}^{\lfloor \frac{j}{2} \rfloor} I\left( \left\{1,f_4\right\}^k, 1, f_3, \left\{f_2\right\}^{j-2k}; q_2\right)
 \right\}.
\end{align}
This is one of the main results of this paper.
Eq.~(\ref{main_result}) gives us easily the $j$-th term in the $\eps$-expansion of $S_{111}(2-2\eps,t)$
in terms of iterated integrals of modular forms: One expands the exponential function and collects
all terms contributing to the $j$-th power of $\eps$. Products of iterated integrals can be converted
to a sum of single iterated integrals with the help of the shuffle product.
We would like to point that the prefactor $\psi_1/\pi$ is itself a modular form 
($\psi_1/\pi \in \mathcal{M}_1(12,\chi_1)$).
Thus the structure of our result is
\begin{align}
 \mbox{sunrise} =
 \mbox{modular form}
 \times
 \mbox{iterated integrals of modular forms}.
\end{align}
The key point to obtain this solution was the transformation from the original integral $S_{111}(2-2\eps,t)$
to the integral $\tilde{E}_{111}(2-2\eps,q_2)$.
The latter satisfies the rather simple differential equation given in eq.~(\ref{diff_eq_Etilde}).
The differential operator on the left-hand side is a quadratic polynomial in $\eps$.
Let us briefly comment on the consequences if we transform to a new basis integral, such that the corresponding
differential operator is linear in $\eps$.
We recall that $f_4=f_1^4$.
The transformation
\begin{align}
 \tilde{E}_{111}'(2-2\eps,q_2)
 & =
 e^{-\eps I(f_1^2;q_2)}
 \tilde{E}_{111}(2-2\eps,q_2)
\end{align}
eliminates the $\eps^2$-term of the differential operator and we find as differential equation for 
$\tilde{E}_{111}'$
\begin{align}
\label{diff_eq_Etildeprime}
 \left[
  \left( q_2 \frac{d}{dq_2} \right)^2 + 2 \eps f_1^2 \left( q_2 \frac{d}{dq_2} \right)
 + 2 \eps f_1 \left( q_2 \frac{d f_1}{dq_2} \right) \right] \tilde{E}_{111}'\left(2-2\eps,q_2\right)
 & = 
 f_3 e^{\eps I(f_2-f_1^2;q_2)}.
\end{align}
The differential operator is now linear in $\eps$, however we left the space of modular forms:
The derivative of a modular form (e.g. $df_1/d\ln q_2$) is in general not a modular form again,
just a quasi-modular form \cite{Kaneko:1995aa,Matthes:2017aa}.


\section{The elliptic curve related to the maximal cut}
\label{sect:Gamma_6}

In expressing the sunrise integral and the kite integral to all orders in the dimensional regularisation parameter $\eps$
as iterated integrals of modular forms we worked up to now with modular forms of $\Gamma_1(12)$.
Bloch and Vanhove gave in \cite{Bloch:2013tra} an expression for the first term in the $\eps$-expansion of the sunrise integral
related to the congruence subgroup $\Gamma_1(6)$.
This raises the question if we may express the sunrise integral and the kite integral to all orders 
in the dimensional regularisation parameter $\eps$ as iterated integrals of modular forms 
from the smaller space $\mathcal{M}_k(\Gamma_1(6))$ only, 
instead of the larger space $\mathcal{M}_k(\Gamma_1(12))$.
The answer is yes.
One possibility which leads us with a standard choice of periods directly to $\Gamma_1(6)$ starts from an elliptic curve
related to the maximal cut of the sunrise integral.
The maximal cut is interesting in its own right \cite{Primo:2016ebd} 
and we discuss the calculation based on an elliptic curve obtained from the maximal cut in this section.
However, there is a small trade-off:
In the region $t<0$ the sunrise integral and the kite integral are real.
The previously defined nome $q_2$ related to the elliptic curve obtained from the second graph polynomial
is real as well in this region.
Furthermore, the nome $q_2$ when viewed as a function of $t$ is smooth around $t=0$.
These two properties do no longer hold for the elliptic curve obtained from the maximal cut.
Let us call the nome in the maximal cut-case $q_{2,C}$, it will be defined below.
One finds that $q_{2,C}$ is complex for $t<0$ and 
$q_{2,C}(t)$ is continuous, but not differentiable at $t=0$.
There is nothing wrong with the fact that $q_{2,C}(t)$ is not differentiable at $t=0$.
The point $t=0$ is a regular singular point of the differential equation.
However, the sunrise integral and the kite integral are regular at this point and one might prefer working with variables
which are smooth around $t=0$.
This simplifies the discussion of the analytic continuation of the Feynman integrals \cite{Bogner:2017vim}.
Geometrically we have the following situation: The elliptic curve associated to the second graph polynomial
defines for $t<0$ a rectangular lattice.
This is not the case for the elliptic curve associated to the maximal cut.
The final results of the two cases are closely related. If we set $q_C=q_{2,C}^2$ the results
are related by the simple substitution $q_C=-q_2$.
The function $q_C(t)$ is again differentiable at $t=0$.

Bloch and Vanhove obtained their result by realising that the relevant family of elliptic curves has been studied in the mathematical
literature \cite{Beauville:1982,Stienstra:1985,Maier:2006aa}.
Here we follow a different path: We may obtain the relevant modular forms of $\Gamma_1(6)$ by looking at the maximal cuts of the 
sunrise integral.
There are two possibilities to obtain an elliptic curve from the sunrise integral.
The first possibility is the one followed in the main text of this paper:
The zero set of the second graph polynomial
\bq
 {\mathcal F} & = & 0
\eq
defines together with the choice of a rational point as origin an elliptic curve.
Let us denote this curve by $E$.
The periods $\psi_1$ and $\psi_2$ and the nome $q_2$ are then defined as given in section~\ref{sec:EllipticCurve}, following the conventions
of \cite{Carlson}.
The curve $E$ has the $j$-invariant
\bq
\label{j_invariant_Feynman}
 j\left(E\right)
 & = &
 \frac{\left(3m^2+t\right)^3 \left(3 m^6 - 3 m^4 t + 9 m^2 t^2 - t^3\right)^3}
      {m^{12} t^2 \left(m^2-t\right)^3 \left(9m^2-t\right)}.
\eq
There is a second possibility to obtain an elliptic curve.
We look at the maximal cut of the sunrise integral around two space-time dimensions.
The maximal cut is given by
\bq
\label{maxcut}
\lefteqn{
 \mathrm{MaxCut}_{\mathcal C} \; S_{111}\left(2-2\eps,t\right)
 = } & &
 \\
 & &
 \frac{u \mu^2}{\pi^2}
 \int\limits_{\mathcal C} 
 \frac{dP}{\left(P -t \right)^{\frac{1}{2}} \left(P - t + 4 m^2 \right)^{\frac{1}{2}} \left(P^2 + 2 m^2 P - 4 m^2 t + m^4\right)^{\frac{1}{2}}}
 +
 {\mathcal O}\left(\eps\right),
 \nonumber
\eq
where $u$ is an (irrelevant) phase and ${\mathcal C}$ an integration contour.
The denominator of the integrand defines a second elliptic curve, which we denote by $E_C$:
\bq
\label{E_73_maxcut}
 E_{C}
 & : &
 w^2 - \left(z - \frac{t}{\mu^2} \right) 
       \left(z - \frac{t - 4 m^2}{\mu^2} \right) 
       \left(z^2 + \frac{2 m^2}{\mu^2} z + \frac{m^4 - 4 m^2 t}{\mu^4} \right)
 \; = \; 0.
\eq
The curve $E_C$ has the $j$-invariant
\bq
\label{j_invariant_maxcut}
 j\left(E_{C}\right)
 & = &
 \frac{\left(3m^2+t\right)^3 \left(3 m^6 + 75 m^4 t - 15 m^2 t^2 + t^3\right)^3}
      {m^6 t \left(m^2-t\right)^6 \left(9m^2-t\right)^2}.
\eq
The $j$-invariants of eq.~(\ref{j_invariant_Feynman}) and eq.~(\ref{j_invariant_maxcut}) differ, 
therefore the two elliptic curves $E$ and $E_{C}$
are not related by a modular $\mathrm{PSL}(2,{\mathbb Z})$-transformation.
They are however related by a quadratic transformation.
To see this, let us denote the roots of the quartic polynomial in eq.~(\ref{E_73_maxcut})
by
\bq
 z_1 \; = \; \frac{t-4m^2}{\mu^2},
 \;\;\;
 z_2 \; = \; \frac{-m^2-2m\sqrt{t}}{\mu^2},
 \;\;\;
 z_3 \; = \; \frac{-m^2+2m\sqrt{t}}{\mu^2},
 \;\;\;
 z_4 \; = \; \frac{t}{\mu^2}.
\eq
We consider a neighbourhood of $t=0$ without the branch cut of $\sqrt{t}$ along the negative real axis.
The correct physical value is specified by Feynman's $i\delta$-prescription: $t\rightarrow t+i\delta$.
We further set
\bq
 k_{C}^2 
 \; = \;
 \frac{\left(z_3-z_2\right)\left(z_4-z_1\right)}{\left(z_3-z_1\right)\left(z_4-z_2\right)},
 & &
 k_{C}'{}^2
 \; = \;
 \frac{\left(z_2-z_1\right)\left(z_4-z_3\right)}{\left(z_3-z_1\right)\left(z_4-z_2\right)}.
\eq
A standard choice of periods is then
\bq
 \psi_{1,C}
 \; = \;
 \frac{4 \mu^2 K\left( k_{C} \right)}{\left(m+\sqrt{t}\right)^{\frac{3}{2}} \left(3m-\sqrt{t}\right)^{\frac{1}{2}}},
 & &
 \psi_{2,C}
 \; = \;
 \frac{4i \mu^2 K\left( k_{C}' \right)}{\left(m+\sqrt{t}\right)^{\frac{3}{2}} \left(3m-\sqrt{t}\right)^{\frac{1}{2}}}.
\eq
We denote the ratio of the two periods and the nome by
\bq
 \tau_{C}
 \;\; = \;\;
 \frac{\psi_{2,C}}{\psi_{1,C}},
 & &
 q_{2,C} \;\; = \;\; e^{i \pi \tau_C}.
\eq
Comparing $(\psi_{1,C},\psi_{2,C})$ with $(\psi_{1},\psi_{2})$ we find
in a neighbourhood of $t=0$
\bq
 \psi_{1,C} \; = \; \psi_{1},
 & &
 2 \psi_{2,C} \; = \; \psi_{2}+\psi_{1}.
\eq
For the Wronskian we have
\bq
\label{Wronskian_relation}
 W_C 
 & = & 
 \psi_{1,C} \frac{d}{dt} \psi_{2,C} - \psi_{2,C} \frac{d}{dt} \psi_{1,C}
 \;\; = \;\;
 - \frac{6 \pi i \mu^4}{t\left(t-m^2\right)\left(t-9m^2\right)}
 \;\; = \;\;
 \frac{1}{2} W.
\eq
Thus, the lattice generated by the periods $(\psi_{1},\psi_{2})$ is a sub-lattice of the one generated by the periods $(\psi_{1,C},\psi_{2,C})$.
\begin{figure}
\begin{center}
\includegraphics[scale=1.0]{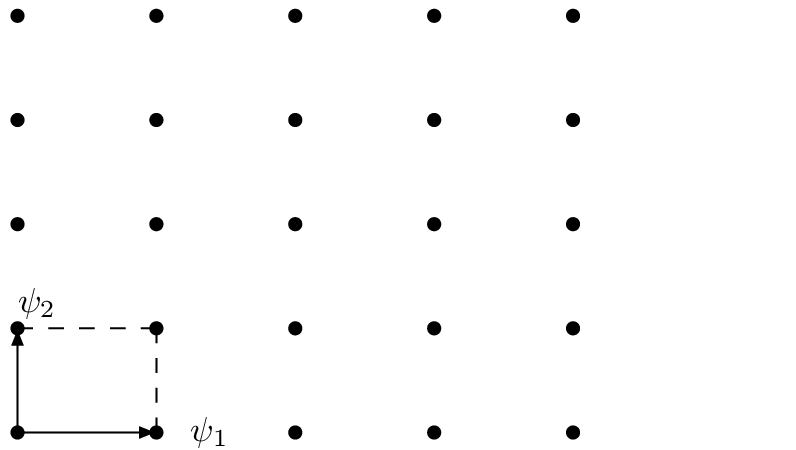}
\includegraphics[scale=1.0]{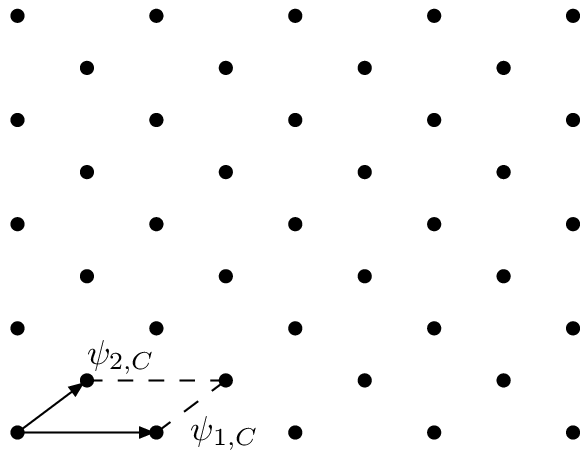}
\end{center}
\caption{
The lattices generated by the periods $(\psi_{1},\psi_{2})$ (left) and $(\psi_{1,C},\psi_{2,C})$ (right).
The lattice generated by $(\psi_{1},\psi_{2})$ is rectangular for $t<0$.
It is a sub-lattice of the one generated by $(\psi_{1,C},\psi_{2,C})$.
}
\label{fig_lattice}
\end{figure}
This is illustrated in fig.~(\ref{fig_lattice}).
The quantities
$\tau_{C}$ and $\tau$ are related by
\bq
 \tau_{C}
 \; = \;
 \frac{\tau+1}{2},
 & &
 \tau
 \; = \; 
 2 \tau_{C} - 1,
\eq
and therefore
\bq
 q_{2,C}^2 & = & - q_{2}.
\eq
We note that for $t<0$ the ratio $\tau$ is purely imaginary and hence $q_2$ is real.
On the other hand, the ratio $\tau_C$ has in this region a non-vanishing real part and hence $q_{2,C}$ is complex.
Let us further note that both pairs $(\psi_{1,C},\psi_{2,C})$ and $(\psi_{1},\psi_{2})$
are a pair of two independent solutions of the differential equation
\bq
 \left[ 
    \frac{d^2}{dt^2} 
    + \left( \frac{1}{t} + \frac{1}{t-m^2} + \frac{1}{t-9m^2} \right) \frac{d}{dt} 
    + \frac{1}{m^2} \left( - \frac{1}{3 t} + \frac{1}{4\left(t-m^2\right)} + \frac{1}{12\left(t-9m^2\right)} \right)
 \right] \psi 
 & = & 0.
 \nonumber \\
\eq
This is immediately clear: If $(\psi_{1,C},\psi_{2,C})$ (or $(\psi_{1},\psi_{2})$) is a pair of two independent solutions, so is any other pair
obtained by a $\mathrm{GL}(2,{\mathbb C})$-transformation.

Let us now look at the modular aspects of the elliptic curve $E_C$.
The analogue of eq.~(\ref{modular_function_t}) reads 
\begin{align}
\label{hauptmodul}
 t & = 
 9 m^2 
 \frac{\eta\left(6\tau_C\right)^8 \eta\left(\tau_C\right)^4}
      {\eta\left(2\tau_C\right)^8 \eta\left(3\tau_C\right)^4}.
\end{align}
This is a modular function of level $6$.
Eq.~(\ref{hauptmodul}) is obtained from expressing $k_C^2 k_C'{}^2$ on the one hand as a function of $t$, and
on the other hand as a function of $\tau_C$
\bq
 16 \frac{m^3 \sqrt{t}\left(3m+\sqrt{t}\right)\left(m-\sqrt{t}\right)^3}{\left(3m-\sqrt{t}\right)^2\left(m+\sqrt{t}\right)^6}
 \;\; = \;\;
 k_C^2 k_C'{}^2
 \;\; = \;\;
 16 \frac{\eta\left(\frac{\tau_C}{2}\right)^{24} \eta\left(2\tau_C\right)^{24}}{\eta\left(\tau_C\right)^{48}},
\eq
and by solving for $t$ as a power series of $q_{2,C}=\exp(i \pi \tau_C)$.
We may now express all integration kernels in terms of modular forms of $\Gamma_1(6)$.
To this aim we denote now with a slight abuse of notation by
$\chi_0$ and $\chi_1$ the characters of $\Gamma_1(6)$ induced by the primitive characters
\begin{align}
\bar{\chi}_0 = \left( \dfrac{1}{n} \right), \qquad
\bar{\chi}_1 = \left( \dfrac{-3}{n} \right).
\end{align}
We have $\psi_{1,C}/\pi \in \mathcal{M}_1(6,\chi_1)$:
\begin{align}
 \frac{\psi_1}{\pi} 
 & =
 \frac{2 \mu^2}{\sqrt{3} m^2}
 \frac{\eta\left(3\tau_C\right) \eta\left(2 \tau_C\right)^6}{\eta\left(\tau_C\right)^{3} \eta\left(6\tau_C\right)^2}
 \nonumber \\
 & =
 \frac{6 \mu^2}{\sqrt{3} m^2}
 \left[ 
 E_1\left(\tau_C;\bar{\chi}_0,\bar{\chi}_1\right)
 + E_1\left(2\tau_C;\bar{\chi}_0,\bar{\chi}_1\right)
 \right].
\end{align}
In the sunrise integral the integration kernels are now expressed in terms of 
three modular forms $f_{1,C} \in \mathcal{M}_1(6,\chi_1)$,
$f_{2,C} \in \mathcal{M}_2(6,\chi_0)$ and
$f_{3,C} \in \mathcal{M}_3(6,\chi_1)$, with
\begin{align}
 f_{1,C} 
 & =
 \frac{\left(t+3m^2\right)}{2 \sqrt{6} \mu^2} \; \frac{\psi_{1,C}}{\pi}
 \; = \;
 3 \sqrt{2} E_1(\tau_C,\bar{\chi}_0,\bar{\chi}_1),
 \nonumber \\
 f_{2,C} 
 & =
 \frac{1}{2 i \pi} \frac{\psi_{1,C}^2}{W_C} \frac{\left(3t^2-10m^2 t - 9 m^4 \right)}{2 t \left(t-m^2\right) \left(t-9m^2\right)}
 \; = \;
 -10 B_{2,2}(\tau_C) + 4 B_{2,3}(\tau_C) - 2 B_{2,6}(\tau_C),
 \nonumber \\
 f_{3,C} 
 & =
 \frac{\mu^2 \psi_{1,C}^3}{4 \pi W_C^2}
 \;
 \frac{6}{t \left(t-m^2\right)\left(t-9m^2\right)}
 \; = \;
 -3 \sqrt{3} 
 \frac{ \eta\left(\tau_C\right)^{5} \eta\left(3\tau_C\right) \eta\left(6\tau_C\right)^4 }
      { \eta\left(2\tau_C\right)^{4} }
 \nonumber \\
 & =
 -3 \sqrt{3} \left[
 E_3\left(\tau_C;\bar{\chi}_1,\bar{\chi}_0\right)
 - 8 E_3\left(2\tau_C;\bar{\chi}_1,\bar{\chi}_0\right)
 \right].
\end{align}
We set further $f_{4,C} = f_{1,C}^4$.

In the kite integral we have in addition three modular forms $g_{2,0,C}, g_{2,1,C}, g_{2,9,C} \in \mathcal{M}_2(6,\chi_0)$ 
and two modular forms $g_{3,0,C}, g_{3,1,C} \in \mathcal{M}_3(6,\chi_1)$.
The weight $2$ modular forms are
\begin{align}
 g_{2,0,C}
 & =
 \frac{1}{2 i \pi} \frac{\psi_{1,C}^2}{W_C} \frac{1}{t}
 \; =  \;
 \frac{\eta\left(\tau_C\right)^{4}\eta\left(3\tau_C\right)^{4}}{\eta\left(2\tau_C\right)^{2}\eta\left(6\tau_C\right)^{2}}
 \; = \;
 - 8 B_{2,2}(\tau_C) - 4 B_{2,3}(\tau_C) + 8 B_{2,6}(\tau_C),
 \nonumber \\
 g_{2,1,C}
 & =
 \frac{1}{2 i \pi} \frac{\psi_{1,C}^2}{W_C} \frac{1}{t-m^2}
 \; = \;
 -9
 \frac{\eta\left(3\tau_C\right)^3\eta\left(6\tau_C\right)^3}{\eta\left(\tau_C\right)\eta\left(2\tau_C\right)}
 \; = \;
 - 9 B_{2,2}(\tau_C) - 3 B_{2,3}(\tau_C) + 3 B_{2,6}(\tau_C),
 \nonumber \\
 g_{2,9,C}
 & =
 \frac{1}{2 i \pi} \frac{\psi_{1,C}^2}{W_C} \frac{1}{t-9m^2}
 \; = \;
 -
 \frac{\eta\left(\tau_C\right)^{7}\eta\left(6\tau_C\right)^{7}}{\eta\left(2\tau_C\right)^{5}\eta\left(3\tau_C\right)^{5}}
 \; = \;
 - 5 B_{2,2}(\tau_C) + 5 B_{2,3}(\tau_C) - B_{2,6}(\tau_C).
\end{align}
The weight $3$ modular forms are given by
\begin{align}
 g_{3,0,C}
 & =
 \frac{1}{2 i \pi \mu^2} \frac{\psi_{1,C}^2}{W_C} \frac{\psi_{1,C}}{\pi}
 \; = \;
 6 \sqrt{3} 
 \frac{ \eta\left(\tau_C\right)^{5} \eta\left(3\tau_C\right) \eta\left(6\tau_C\right)^{4} }
      { \eta\left(2\tau_C\right)^{4} }
 \nonumber \\
 & =
 6 \sqrt{3} \left[
 E_3\left(\tau_C;\bar{\chi}_1,\bar{\chi}_0\right)
 - 8 E_3\left(2\tau_C;\bar{\chi}_1,\bar{\chi}_0\right)
 \right],
 \nonumber \\
 g_{3,1,C}
 & =
 \frac{1}{2 i \pi \mu^2} \frac{\psi_{1,C}^2}{W_C} \frac{\psi_{1,C}}{\pi} \frac{t}{t-m^2}
 \; = \;
 - 54 \sqrt{3} \frac{\eta\left(6\tau_C\right)^{9}}{\eta\left(2\tau_C\right)^{3}}
 \; = \;
 -54 \sqrt{3} E_3\left(2\tau_C;\bar{\chi}_1,\bar{\chi}_0\right).
\end{align}
According to a theorem by Sebbar \cite{Sebbar:2002aa}
the space of modular forms for $\Gamma_1(6)$
is a polynomial ring\footnote{We thank the referee for pointing this out.} with two generators of weight $1$.
As generators we may take
\bq
 e_1 \; = \; E_1\left(\tau_C;\bar{\chi}_0,\bar{\chi}_1\right),
 & &
 e_2 \; = \; E_1\left(2\tau_C;\bar{\chi}_0,\bar{\chi}_1\right).
\eq
The theorem implies that we should be able to express all occurring modular forms of $\Gamma_1(6)$ as polynomials
in the two generators.
Indeed we find
\bq
 \frac{\psi_1}{\pi} 
 & = &
 2 \sqrt{3}
 \frac{\mu^2}{m^2} \left( e_1 + e_2 \right),
 \nonumber \\
 f_{1,C}
 & = &
 3 \sqrt{2} e_1,
 \nonumber \\
 f_{2,C} 
 & = &
 -6 \left( e_1^2 + 6 e_1 e_2 - 4 e_2^2 \right),
 \nonumber \\
 f_{3,C} 
 & = &
 36 \sqrt{3} \left( e_1^3 - e_1^2 e_2 - 4 e_1 e_2^2 + 4e_2^3 \right),
 \nonumber \\
 g_{2,0,C} 
 & = &
 - 12 \left( e_1^2 - 4 e_2^2 \right),
 \nonumber \\
 g_{2,1,C} 
 & = &
 - 18 \left( e_1^2 + e_1 e_2 - 2 e_2^2 \right),
 \nonumber \\
 g_{2,9,C} 
 & = &
 6 \left( e_1^2 - 3 e_1 e_2 + 2 e_2^2 \right),
 \nonumber \\
 g_{3,0,C} 
 & = &
 -72 \sqrt{3} \left( e_1^3 - e_1^2 e_2 - 4 e_1 e_2^2 + 4e_2^3 \right),
 \nonumber \\
 g_{3,1,C} 
 & = &
 - 108 \sqrt{3} \left( e_1^3 - 3 e_1 e_2^2 + 2 e_2^3 \right).
\eq
The relation with the previously defined modular forms of $\Gamma_1(12)$ is given by
\bq
 f_i\left(\tau_2\right) \;\; = \;\; f_{i,C}\left(\tau_C\right),
 & &
 g_{i,j}\left(\tau_2\right) \;\; = \;\; g_{i,j,C}\left(\tau_C\right),
\eq
where $\tau_C$ is related to $\tau_2$ by $\tau_C = \tau_2 + 1/2$.
Let us introduce
$q_C=q_{2,C}^2=\exp(2 i \pi \tau_C)=-q_2$.
The all-order expression for the sunrise integral in terms of iterated integrals of modular forms of $\Gamma_1(6)$ reads
\begin{align}
 S_{111}\left(2-2\eps,t\right)
 &= 
 \frac{\psi_{1,C}}{\pi}
 e^{-\eps I(f_{2,C};q_{C}) -2\eps L -2\gamma_E \eps + 2\sum\limits_{n=2}^\infty \frac{\left(-1\right)^n}{n} \zeta_n \eps^n} 
 \nonumber \\
 &
 \left\{
 \left[
 \sum\limits_{j=0}^\infty 
 \left(
 \eps^{2j} I\left(\left\{1,f_{4,C}\right\}^j;q_{C} \right)
 -
 \frac{1}{2}
 \eps^{2j+1} I\left(\left\{1,f_{4,C}\right\}^j,1;q_{C} \right)
 \right)
 \right]
 \sum\limits_{k=0}^\infty \eps^k B^{(k)}\left(2,0\right)
 \right. \nonumber \\
 & 
 \left.
 +
  \sum\limits_{j=0}^\infty 
   \eps^j 
   \sum\limits_{k=0}^{\lfloor \frac{j}{2} \rfloor} I\left( \left\{1,f_{4,C}\right\}^k, 1, f_{3,C}, \left\{f_{2,C}\right\}^{j-2k}; q_{C}\right)
 \right\}.
\end{align}
The boundary values $B^{(k)}(2,0)$ are the ones from eq.~(\ref{def_boundary_values}).


\section{The choice of the periods}
\label{sect:choice_periods}

Let us return to the case, where the elliptic curve is defined by the second graph polynomial
\bq
 {\mathcal F} & = & 0.
\eq
We defined two periods $\psi_1$ and $\psi_2$ in eq.~(\ref{def_periods}).
This choice was motivated by the fact, that the lattice generated by the periods is rectangular in the Euclidean region, and we chose
the periods such that $\psi_1$ is real and $\psi_2$ is purely imaginary in the Euclidean region.
Any other choice of periods is related to the original one by a 
$\text{SL}_2(\mathbb{Z})$-transformation:
\bq
 \left(\begin{array}{c}
 \psi_2' \\
 \psi_1' \\
 \end{array} \right)
 & = &
 \left( \begin{array}{cc} 
  a & b \\
  c & d \\
 \end{array} \right)
 \left(\begin{array}{c}
 \psi_2 \\
 \psi_1 \\
 \end{array} \right),
 \;\;\;\;\;\;\;\;\;
 \gamma \; = \; 
 \left( \begin{array}{cc} 
  a & b \\
  c & d \\
 \end{array} \right)
 \; \in \; 
 \text{SL}_2(\mathbb{Z}).
\eq
This induces a transformation on $\tau$:
\bq
 \tau' & = & \gamma\left(\tau\right) \; = \; \frac{a\tau + b}{c \tau + d}.
\eq
In this section we discuss the effects related to the choice of periods.
Let us start with an example:
The transformation
\bq
 T
 & = &
 \left( \begin{array}{cc} 
  1 & 1 \\
  0 & 1 \\
 \end{array} \right)
\eq
induces 
\bq
 \tau' \; = \; \tau + 1,
 & \mbox{and} &
 q_2' \; = \; - q_2.
\eq
An inspection of the results from section~\ref{sect:Gamma_6} shows that the transformation $q_2'=-q_2$ 
has the effect that all occurring transformed modular forms belong to the smaller space ${\mathcal M}_k(\Gamma_1(6))$.

Note that there is no contradiction with the fact that $T \in \Gamma_1(12)$.
In section~\ref{sec:integration_kernels} we showed that the integration kernels are modular forms of $\Gamma_1(12)$
in the variable $\tau_2=\tau/2$.
Here, we discuss the effects of $\text{SL}_2(\mathbb{Z})$-transformations on the choice of periods.
The transformation $\tau'=\tau+1$ corresponds to $\tau_2' = \tau_2 + 1/2$ or $\tau_2'=T_2(\tau_2)$ with
\bq
 T_2
 & = &
 \left( \begin{array}{cc} 
  1 & \frac{1}{2} \\
  0 & 1 \\
 \end{array} \right).
\eq
The matrix $T_2$ belongs to $\text{SL}_2(\mathbb{Q})$, but not to $\text{SL}_2(\mathbb{Z})$.
If on the other hand we consider $\tau_2''=T(\tau_2)=\tau_2+1$, we have $\tau''=\tau+2$ and $q_2''=q_2$.
This shows that any modular form $f(\tau_2) \in \Gamma_1(N)$ is invariant under $\tau_2''=T(\tau_2)$, as it should.

Let us now study the issue more systematically:
We consider $ t \in \hat{\mathbb C} = {\mathbb C} \cup \{\infty\}$. A choice of periods $\psi_1'$, $\psi_2'$ defines
a map 
\bq
 \hat{\mathbb C} & \rightarrow & {\mathbb H}^\ast,
 \nonumber \\
 t & \rightarrow & \tau' \; = \; \frac{\psi_2'}{\psi_1'} \; = \; \gamma\left(\frac{\psi_2}{\psi_1}\right).
\eq
We are interested in inverting this map in a neighbourhood of a point $t_0 \in \hat{\mathbb C}$
with $\tau'(t_0) = i \infty$.
Of particular interest is the case, where $t_0$ is one of the singular points of the differential equation:
$t_0 \in \{0, m^2, 9m^2, \infty\}$.
In section~\ref{sec:integration_kernels} we used the modular lambda function for this purpose.
The modular lambda function is a modular function for $\Gamma(2)$.
The congruence subgroup $\Gamma(2)$ has index $6$ in $\text{SL}_2(\mathbb{Z})$.
Thus any $\gamma \in \text{SL}_2(\mathbb{Z})$ belongs to one of the six $\Gamma(2)$-cosets in $\text{SL}_2(\mathbb{Z})$.
The congruence subgroup $\Gamma(2)$ is normal in $\text{SL}_2(\mathbb{Z})$, therefore the left and right cosets are identical.
For any coset one finds a representative $\gamma_{\mathrm{rep}}$ and a $t_0 \in \{0,m^2,9m^2,\infty\}$ such that 
$\tau'(t_0) = i \infty$.
\begin{table}
\begin{center}
\begin{tabular}{c|c|ccc|c|c|l}
 $\gamma_{\mathrm{rep}}$ & $\tau'$ & $e_1'$ & $e_2'$ & $e_3'$ & $\lambda(\tau')=\frac{e_3'-e_2'}{e_1'-e_2'}$ & $t_0$ & relation \\ 
 & & & & & & & \\
\hline
 & & & & & & & \\
 $\left(\begin{array}{rr} 1 & 0 \\ 0 & 1 \\ \end{array} \right)$ & $\tau$ & $e_1$ & $e_2$ & $e_3$ & $\lambda = \frac{e_3-e_2}{e_1-e_2}$ & $0$ &
 $t = 
 -9 m^2 
 \frac{\eta\left(2\tau_2'\right)^4 \eta\left(3\tau_2'\right)^4 \eta\left(12 \tau_2'\right)^4}
      {\eta\left(\tau_2'\right)^4 \eta\left(4 \tau_2'\right)^4 \eta\left(6 \tau_2'\right)^4}$
\\
 & & & & & & & \\
 $\left(\begin{array}{rr} 1 & 1 \\ 0 & 1 \\ \end{array} \right)$ & $\tau+1$ & $e_1$ & $e_3$ & $e_2$ & $-\frac{\lambda}{1-\lambda} = \frac{e_2-e_3}{e_1-e_3}$ & $0$ &
 $ t = 
 9 m^2 
 \frac{\eta\left(\tau_2'\right)^4 \eta\left(6 \tau_2'\right)^8}
      {\eta\left(3 \tau_2'\right)^4 \eta\left(2 \tau_2'\right)^8}$ \\
 & & & & & & & \\
 $\left(\begin{array}{rr} 0 & -1 \\ 1 & 0 \\ \end{array} \right)$ & $-\frac{1}{\tau}$ & $e_2$ & $e_1$ & $e_3$ & $1-\lambda = \frac{e_3-e_1}{e_2-e_1}$ & $\infty$ &
 $ -\frac{1}{t} = 
 \frac{1}{m^2}
 \frac{\eta\left(2\tau_6'\right)^4 \eta\left(3 \tau_6'\right)^4 \eta\left(12 \tau_6'\right)^4}
      {\eta\left(\tau_6'\right)^4 \eta\left(4 \tau_6'\right)^4 \eta\left(6 \tau_6'\right)^4}$ \\
 & & & & & & & \\
 $\left(\begin{array}{rr} 3 & -1 \\ 1 & 0 \\ \end{array} \right)$ & $\frac{3\tau-1}{\tau}$ & $e_2$ & $e_3$ & $e_1$ & $-\frac{1-\lambda}{\lambda} = \frac{e_1-e_3}{e_2-e_3}$ & $\infty$ &
 $ - \frac{1}{t} = 
 - \frac{1}{m^2}
 \frac{\eta\left(\tau_6'\right)^4 \eta\left(6 \tau_6'\right)^8}
      {\eta\left(3 \tau_6'\right)^4 \eta\left(2 \tau_6'\right)^8}$ \\
 & & & & & & & \\
 $\left(\begin{array}{rr} 3 & 2 \\ 1 & 1 \\ \end{array} \right)$ & $\frac{3\tau+2}{\tau+1}$ & $e_3$ & $e_2$ & $e_1$ & $\frac{1}{\lambda} = \frac{e_1-e_2}{e_3-e_2}$ & $m^2$ &
 $t-m^2 = 
 - 8 m^2
 \frac{\eta\left(\tau_3'\right)^3 \eta\left(6 \tau_3'\right)^9}
      {\eta\left(2 \tau_3'\right)^3 \eta\left(3 \tau_3'\right)^9}$ \\
 & & & & & & & \\
 $\left(\begin{array}{rr} 0 & -1 \\ 1 & 1 \\ \end{array} \right)$ & $-\frac{1}{\tau+1}$ & $e_3$ & $e_1$ & $e_2$ & $\frac{1}{1-\lambda} = \frac{e_2-e_1}{e_3-e_1}$ & $m^2$ &
  $t-m^2 = 
 - 8 m^2
 \frac{\eta\left(\tau_3'\right)^3 \eta\left(6 \tau_3'\right)^9}
      {\eta\left(2 \tau_3'\right)^3 \eta\left(3 \tau_3'\right)^9}$ \\
\end{tabular}
\end{center}
\caption{
\it The table shows in the first column representatives for the six $\Gamma(2)$-cosets in $\text{SL}_2(\mathbb{Z})$.
The second column expresses $\tau'$ in terms of $\tau$.
The third column gives the permutation of the roots appearing in the Weierstrass normal form.
The fourth column expresses $\lambda(\tau')$ in terms of $\lambda(\tau)$.
The fifth column gives the point $t_0$ such that $\tau'(t_0)=i\infty$.
Finally, the last column relates the variable $t$ to an eta quotient in $\tau'$.
Here the notation $\tau_N'=\tau'/N$ is used.
}
\label{table_choice_periods}
\end{table}
For these representatives $\gamma_{\mathrm{rep}}$ we may invert the map $t \rightarrow \tau'$ in a neighbourhood of $t=t_0$.
The coset representatives $\gamma_{\mathrm{rep}}$, the points $t_0$ and the result of the inversion are tabulated in table~(\ref{table_choice_periods}), along
with additional useful information.

An arbitrary $\gamma \in \text{SL}_2(\mathbb{Z})$ may be written as
\bq
 \gamma & = & \gamma_2 \gamma_{\mathrm{rep}}
\eq
with $\gamma_2 \in \Gamma(2)$ and $\gamma_{\mathrm{rep}}$ being one of the six coset representatives tabulated in table~(\ref{table_choice_periods}).
If the result of the inversion for the case $\tau'=\gamma_{\mathrm{rep}}(\tau)$ is written as $t=F(\tau')$, then we have for the case
$\tau' = \gamma(\tau) = \gamma_2(\gamma_{\mathrm{rep}}(\tau))$
\bq
 t & = & F\left(\gamma_2^{-1}\left(\tau'\right)\right).
\eq
The coset representatives are of course not unique. For example we may also choose for the fifth coset 
the representative
\bq
\label{further_example_representative}
 \gamma & = & 
 \left( \begin{array}{rr}
 1 & 0 \\
 3 & 1 \\
 \end{array} \right)
 \; = \;
 \left( \begin{array}{rr}
 1 & -2 \\
 2 & -3 \\
 \end{array} \right)
 \left( \begin{array}{rr}
 3 & 2 \\
 1 & 1 \\
 \end{array} \right).
\eq
For this $\gamma$ we have $\tau'(9 m^2) = i \infty$. Inverting this function in a neighbourhood of $t_0=9m^2$
one finds
\bq
 t - 9 m^2
 & = &
 72 m^2
 \frac{\eta\left(2\tau'\right) \eta\left(6 \tau'\right)^5}
      {\eta\left(3 \tau'\right) \eta\left(\tau'\right)^5},
\eq
which is the relation given in ref.~\cite{Bloch:2013tra}.
The transformed periods $\psi_2'$ and $\psi_1'$ generate the same lattice as $\psi_2$ and $\psi_1$.
The transformation in eq.~(\ref{further_example_representative}) induces
\bq
 \tau' & = & \frac{\tau}{3\tau+1}.
\eq
Note however that the relation between $\tau'$ and $\tau_2$ is
\bq
 \tau' & = & \frac{2\tau_2}{6 \tau_2 +1},
\eq
which is not a $\text{SL}_2(\mathbb{Z})$-transformation (and there is no reason 
for this transformation to be a $\text{SL}_2(\mathbb{Z})$-transformation).


\section{Conclusions}
\label{sect:conclusions}

In this paper we showed that certain Feynman integrals, which cannot be expressed in terms of multiple polylogarithms,
have a representation as a linear combinations of iterated integrals of modular forms.
Specifically, we showed that equal mass sunrise integral and the kite integral can expressed to all orders 
in the dimensional regularisation parameter $\varepsilon$ in this form.
This structure is very appealing, as it generalises the pattern observed for Feynman integrals evaluating to multiple polylogarithms
to the elliptic setting.
For the former we have the structure
\begin{align}
 \mbox{algebraic prefactor}
 \times
 \mbox{iterated integrals of rational/algebraic functions},
\end{align}
for the latter we find
\begin{align}
 \mbox{modular form}
 \times
 \mbox{iterated integrals of modular forms}.
\end{align}
In addition we presented a compact formula, expressing the two-loop sunrise integral to all orders in $\varepsilon$
as iterated integrals of modular forms.

In our calculations we started from an elliptic curve, which we may either obtain from the Feynman parameter representation
or from the maximal cut.
These two elliptic curves are not identical, however they are related by a quadratic transformation.
Of course, both approaches yield the correct result.

The space of modular forms for the integration kernels depends on the elliptic curve we start with and on the choice
of periods for this elliptic curve.
We discussed both aspects.
With a standard choice of periods we find that the integration kernels belong to
$\mathcal{M}_k(\Gamma_1(12))$ in the case where the elliptic curve is obtained from the 
Feynman parameter representation, and to
$\mathcal{M}_k(\Gamma_1(6))$ in the case where the elliptic curve is obtained from the 
maximal cuts.
In the Feynman parameter case, the transformation $\tau'=\tau+1$ transforms the integration kernels
from the larger space $\mathcal{M}_k(\Gamma_1(12))$ into the smaller space $\mathcal{M}_k(\Gamma_1(6))$.

\subsection*{Acknowledgements}

L.A. is grateful for financial support from the research training group GRK 1581.
We would like to thank Stefan M\"uller-Stach for useful discussions,
furthermore S.W. would like to thank Johannes Bl\"umlein for useful discussions.


\begin{appendix}

\ifthenelse{\boolean{arxiveversion}}
{
\section{The Kronecker symbol}
\label{sec:Kronecker_symbol}

Let $a$ be an integer and $n$ a non-zero integer with prime factorisation
$n = u p_1^{\alpha_1} p_2^{\alpha_2} ... p_k^{\alpha_k}$,
where $u \in \{1,-1\}$ is a unit.
The Kronecker symbol is defined by
\bq
 \left( \frac{a}{n} \right)
 & = & 
 \left( \frac{a}{u} \right)
 \left( \frac{a}{p_1} \right)^{\alpha_1}
 \left( \frac{a}{p_2} \right)^{\alpha_2}
 ...
 \left( \frac{a}{p_k} \right)^{\alpha_k}.
\eq
The individual factors are defined as follows:
For a unit $u$ we define
\bq
 \left( \frac{a}{u} \right)
 & = &
 \left\{ \begin{array}{rl}
 1, & u=1, \\
 1, & u=-1, \; a \ge 0, \\
 -1, & u=-1, \; a<0. \\
 \end{array} \right.
\eq
For $p=2$ we define
\bq
 \left( \frac{a}{2} \right)
 & = &
 \left\{ \begin{array}{rl}
 1, & a \equiv \pm 1 \mod 8, \\
 -1, & a \equiv \pm 3 \mod 8, \\
 0, & a \;\; \mbox{even}.  \\
 \end{array} \right.
\eq
For an odd prime $p$ we have
\bq
 \left( \frac{a}{p} \right)
 & = & 
 a^{\frac{p-1}{2}} \mod p
 \;\; = \;\;
 \left\{ \begin{array}{rl}
 1, & a \equiv b^2 \mod p, \\
 -1, & a \not\equiv b^2 \mod p, \\
 0, & a \equiv 0 \mod p. \\
 \end{array} \right.
\eq
We further set
\bq
 \left( \frac{a}{0} \right)
 & = &
 \left\{ \begin{array}{rl}
 1, & a = \pm 1 \\
 0, & \mbox{otherwise}.  \\
 \end{array} \right.
\eq
For any non-zero integer $a$ the mapping 
\bq
 n & \rightarrow & 
 \left( \frac{a}{n} \right)
\eq
is a Dirichlet character.
If $a$ is the discriminant of a quadratic field, then it is a primitive 
Dirichlet character with conductor $|a|$.
One may give a condition for $a$ being the discriminant of a quadratic field \cite{Miyake}.
We first set for $p$ being a prime number, $-1$ or $-2$
\bq
 p^\ast
 & = &
 \left\{ \begin{array}{rl}
 p, & \mbox{if} \quad p \equiv 1 \mod 4, \\
 -p, & \mbox{if} \quad p \equiv -1 \mod 4 \quad \mbox{and} \quad p \neq -1, \\
 -4, & \mbox{if} \quad p = -1, \\
 8, & \mbox{if} \quad p = 2, \\
-8, & \mbox{if} \quad p = -2. \\
 \end{array} \right.
\eq
Then an integer $a$ is the discriminant of a quadratic field if and only if
$a$ is a product of distinct $p^\ast$'s.
} 
{
\section{Generalised Eisenstein series}
\label{sect:eisenstein_space}

The space $\mathcal{M}_k(\Gamma_1(N))$ of modular forms of weight $k$ for the congruence subgroup $\Gamma_1(N)$ decomposes
into a direct sum of spaces of modular forms of weight $k$ for the congruence subgroup $\Gamma_0(N)$ with characters:
\begin{align}
\label{mod_form_space_decomp}
\mathcal{M}_{k}(\Gamma_1(N)) = \bigoplus\limits_{\chi}\ \mathcal{M}_k(N,\chi),
\end{align}
where the sum runs over all Dirichlet characters modulo $N$.
We have similar decompositions for the space of cusp forms and the Eisenstein subspaces:
\begin{align}
\mathcal{S}_{k}(\Gamma_1(N)) = \bigoplus\limits_{\chi}\ \mathcal{S}_k(N,\chi),
 \qquad
\mathcal{E}_{k}(\Gamma_1(N)) = \bigoplus\limits_{\chi}\ \mathcal{E}_k(N,\chi).
\end{align}
A basis for the Eisenstein subspace $\mathcal{E}_k(N,\chi)$ can be given explicitly in terms of generalised Eisenstein series.
Let $\phi$ and $\psi$ be primitive Dirichlet characters with conductors $L$ and $M$, respectively.
We set
\begin{align}
E_{k} (\tau; \phi,\psi) &
 = 
 a_0 + \sum\limits_{m=1}^{\infty} \left( \sum\limits_{d|m} \psi(d) \cdot \phi(m/d) \cdot d^{k-1} \right) q_M^{m},
 \qquad q_M = e^{2\pi i \tau/M}.
\end{align}
The normalisation is such that the coefficient of $q_M$ is one.
The constant term $a_0$ is given by
\begin{align}
 a_0 &
 = 
 \begin{cases}
 -\frac{B_{k,\psi}}{2k}, \qquad &\text{if}\ L=1, \\
0, \qquad &\text{if}\ L>1.
\end{cases}
\end{align}
Note that $L$ denotes the conductor of $\phi$ and the constant term $a_0$ depends therefore on $\phi$ and $\psi$.
The generalised Bernoulli-numbers $B_{k,\psi}$ are defined by
\begin{align}
 \sum\limits_{m=1}^{M} \psi(m) \dfrac{xe^{mx}}{e^{Mx}-1}
 =
 \sum\limits_{k=0}^{\infty} B_{k,\psi} \dfrac{x^k}{k!}. 
\end{align}
The generalised Eisenstein series are modular forms \cite{Stein}:
\begin{Theo}
\label{Eisenstein_series_are_modular}
Suppose $K$ is a positive integer, the Dirichlet characters $\phi$, $\psi$ are as above and $k$ is a positive integer such that
$\phi(-1) \psi(-1) = (-1)^k$.
For $k=1$ we require in addition $\phi(-1)=1$ and $\psi(-1)=-1$.
Except when $k=2$ and $\phi=\psi=1$, the Eisenstein series $E_k(K \tau;\phi,\psi)$ defines an element of
$\mathcal{M}_k(KLM,\tilde{\chi})$, where $\tilde{\chi}$ is the Dirichlet character with modulus $KLM$ induced by $\phi\psi$.
In the case $k=2$, $\phi=\psi=1$ and $K>1$ we use the notation $E_2(\tau)=E_2(\tau;\phi,\psi)$ 
and $B_{2,K}(\tau) = E_2(\tau)- K E_2(K \tau)$.
Then $B_{2,K}(\tau)$ is a modular form in $\mathcal{M}_2(\Gamma_0(K))$.
\end{Theo}
We may now give a basis for the Eisenstein subspace $\mathcal{E}_k(N,\chi)$:
\begin{Theo}
The Eisenstein series in $\mathcal{M}_k(N,\chi)$ coming from theorem~\ref{Eisenstein_series_are_modular} with $KLM | N$ and $\chi$ the Dirichlet character of modulus $N$ induced from $\phi\psi$ form 
a basis for the Eisenstein subspace $\mathcal{E}_k(N,\chi)$.
\end{Theo}
} 

\section{Elliptic generalisations of polylogarithms}
\label{sect:ELi}

In this appendix we collect the definitions of three families of specific functions: These are 
the $\mathrm{ELi}$-functions, the $\overline{\mathrm{E}}$-functions 
and the $\mathrm{E}$-functions \cite{Adams:2015ydq,Adams:2016xah}.
The latter two are just linear combinations of the $\mathrm{ELi}$-functions.
Let us start with the $\mathrm{ELi}$-functions. These are 
functions of $(2l+1)$ variables $x_1$, ..., $x_l$, $y_1$, ..., $y_l$, $q_2$
and $(3l-1)$ indices $n_1$, ..., $n_l$, $m_1$, ..., $m_l$, $o_1$, ..., $o_{l-1}$.
For $l=1$ we set
\begin{align}
 \mathrm{ELi}_{n;m}\left(x;y;q_2\right) 
 & =  
 \sum\limits_{j=1}^\infty \sum\limits_{k=1}^\infty \; \frac{x^j}{j^n} \frac{y^k}{k^m} q_2^{j k}.
\end{align}
For $l>1$ we define
\begin{align}
\lefteqn{
 \mathrm{ELi}_{n_1,...,n_l;m_1,...,m_l;2o_1,...,2o_{l-1}}\left(x_1,...,x_l;y_1,...,y_l;q_2\right) 
 = }
 & \nonumber \\
 & 
 \hspace*{15mm}
 = 
 \sum\limits_{j_1=1}^\infty ... \sum\limits_{j_l=1}^\infty
 \sum\limits_{k_1=1}^\infty ... \sum\limits_{k_l=1}^\infty
 \;\;
 \frac{x_1^{j_1}}{j_1^{n_1}} ... \frac{x_l^{j_l}}{j_l^{n_l}}
 \;\;
 \frac{y_1^{k_1}}{k_1^{m_1}} ... \frac{y_l^{k_l}}{k_l^{m_l}}
 \;\;
 \frac{q_2^{j_1 k_1 + ... + j_l k_l}}{\prod\limits_{i=1}^{l-1} \left(j_i k_i + ... + j_l k_l \right)^{o_i}}.
\end{align}
We have the relations
\begin{align}
\label{ELi_multiplication}
\lefteqn{
 \mathrm{ELi}_{n_1;m_1}\left(x_1;y_1;q_2\right) 
 \mathrm{ELi}_{n_2,...,n_l;m_2,...,m_l;2o_2,...,2o_{l-1}}\left(x_2,...,x_l;y_2,...,y_l;q_2\right) 
 = } & \nonumber \\
 & 
 \hspace*{35mm}
 =
 \mathrm{ELi}_{n_1,n_2,...,n_l;m_1,m_2,...,m_l;0,2o_2,...,2o_{l-1}}\left(x_1,x_2,...,x_l;y_1,y_2,...,y_l;q_2\right) 
\end{align}
and
\begin{align}
\label{ELi_integration}
\lefteqn{
 \int\limits_0^{q_2} \frac{dq_2'}{q_2'}
 \mathrm{ELi}_{n_1,...,n_l;m_1,...,m_l;2o_1,2o_2,...,2o_{l-1}}\left(x_1,...,x_l;y_1,...,y_l;q_2'\right)
 = } & \nonumber \\
 &
 \hspace*{30mm}
 =
 \mathrm{ELi}_{n_1,...,n_l;m_1,...,m_l;2(o_1+1),2o_2,...,2o_{l-1}}\left(x_1,...,x_l;y_1,...,y_l;q_2\right).
\end{align}
It will be convenient to introduce abbreviations for certain linear combinations, which occur quite often.
We define a prefactor $c_n$ and a sign $s_n$, both depending on an index $n$ by
\begin{align}
 c_n = \frac{1}{2} \left[ \left(1+i\right) + \left(1-i\right)\left(-1\right)^n\right] = 
 \left\{ \begin{array}{rl}
 1, & \mbox{$n$ even}, \\
 i, & \mbox{$n$ odd}, \\
 \end{array} \right.
 & &
 s_n = (-1)^n =
 \left\{ \begin{array}{rl}
 1, & \mbox{$n$ even}, \\
 -1, & \mbox{$n$ odd}. \\
 \end{array} \right.
\end{align}
For $l=1$ we define the linear combinations
\begin{align}
 \overline{\mathrm{E}}_{n;m}\left(x;y;q_2\right) 
 & =
 \frac{c_{n+m}}{i}
 \left[
  \mathrm{ELi}_{n;m}\left(x;y;q_2\right)
  - s_{n+m} \mathrm{ELi}_{n;m}\left(x^{-1};y^{-1};q_2\right)
 \right].
\end{align}
More explicitly, we have
\begin{align}
\label{def_Ebar_weight_1}
 \overline{\mathrm{E}}_{n;m}\left(x;y;q_2\right) 
 & =
 \left\{ \begin{array}{ll}
 \frac{1}{i}
 \left[
 \mathrm{ELi}_{n;m}\left(x;y;q_2\right) - \mathrm{ELi}_{n;m}\left(x^{-1};y^{-1};q_2\right)
 \right],
 & \mbox{$n+m$ even,} \\
 & \\
 \mathrm{ELi}_{n;m}\left(x;y;q_2\right) + \mathrm{ELi}_{n;m}\left(x^{-1};y^{-1};q_2\right),
 & \mbox{$n+m$ odd.} \\
 \end{array}
 \right.
\end{align}
For $l>0$ we proceed as follows:
For $o_1=0$ we set
\begin{align}
\label{Ebar_multiplication}
\lefteqn{
 \overline{\mathrm{E}}_{n_1,...,n_l;m_1,...,m_l;0,2o_2,...,2o_{l-1}}\left(x_1,...,x_l;y_1,...,y_l;q_2\right) 
 = } & \nonumber \\
 & 
 \hspace*{30mm}
 =
 \overline{\mathrm{E}}_{n_1;m_1}\left(x_1;y_1;q_2\right) 
 \overline{\mathrm{E}}_{n_2,...,n_l;m_2,...,m_l;2o_2,...,2o_{l-1}}\left(x_2,...,x_l;y_2,...,y_l;q_2\right).
\end{align}
For $o_1 > 0$ we set recursively
\begin{align}
\label{Ebar_integration}
\lefteqn{
 \overline{\mathrm{E}}_{n_1,...,n_l;m_1,...,m_l;2o_1,2o_2,...,2o_{l-1}}\left(x_1,...,x_l;y_1,...,y_l;q_2\right) 
 = 
 } & \nonumber \\
 &
 \hspace*{30mm}
 =
 \int\limits_0^{q_2} \frac{dq_2'}{q_2'} 
 \overline{\mathrm{E}}_{n_1,...,n_l;m_1,...,m_l;2(o_1-1),2o_2,...,2o_{l-1}}\left(x_1,...,x_l;y_1,...,y_l;q_2'\right).
\end{align}
The $\overline{\mathrm{E}}$-functions are
linear combinations of the $\mathrm{ELi}$-functions with the same indices.
More concretely, an $\overline{\mathrm{E}}$-function of depth $l$ can be expressed
as a linear combination of $2^l$ $\mathrm{ELi}$-functions.
We have
\begin{align}
\lefteqn{
 \overline{\mathrm{E}}_{n_1,...,n_l;m_1,...,m_l;2o_1,...,2o_{l-1}}\left(x_1,...,x_l;y_1,...,y_l;q_2\right) 
 = 
 } & \\
 & 
 =
 \sum\limits_{t_1=0}^1 ... \sum\limits_{t_l=0}^1
 \left[ \prod\limits_{j=1}^l \frac{c_{n_j+m_j}}{i} \left( - s_{n_j+m_j} \right)^{t_j} \right]
 \mathrm{ELi}_{n_1,...,n_l;m_1,...,m_l;2o_1,...,2o_{l-1}}\left(x_1^{s_{t_1}},...,x_l^{s_{t_l}};y_1^{s_{t_1}},...,y_l^{s_{t_l}};q_2\right).
 \nonumber
\end{align}
Finally, let us introduce the $\mathrm{E}$-functions. These are closely related to the
$\overline{\mathrm{E}}$-functions, the difference being that the 
$\mathrm{E}$-functions at depth $1$ have a term proportional to $q_2^0$, while the
$\overline{\mathrm{E}}$ do not.
The $\mathrm{E}$-functions are defined as
\begin{align}
\label{relation_E_Ebar}
\lefteqn{
 \mathrm{E}_{n_1,...,n_{l-1},n_l;m_1,...,m_{l-1},m_l;2o_1,...,2o_{l-2},2o_{l-1}}\left(x_1,...,x_{l-1},x_l;y_1,...,y_{l-1},y_l;q_2\right)
  =
} & \nonumber \\
 &
\hspace*{15mm}
 = &&
 \overline{\mathrm{E}}_{n_1,...,n_{l-1},n_l;m_1,...,m_{l-1},m_l;2o_1,...,2o_{l-2},2o_{l-1}}\left(x_1,...,x_{l-1},x_l;y_1,...,y_{l-1},y_l;q_2\right)
 \nonumber \\
 & &&
 +
 \overline{\mathrm{E}}_{n_1,...,n_{l-2},n_{l-1}+o_{l-1};m_1,...,m_{l-2},m_{l-1}+o_{l-1};2o_1,...,2o_{l-2}}\left(x_1,...,x_{l-1};y_1,...,y_{l-1};q_2\right)
 \nonumber \\
 & &&
 \times
 \frac{c_{n_l+m_l}}{2i} \left[ \mathrm{Li}_{n_l}\left( x_l \right) - s_{n_l+m_l} \mathrm{Li}_{n_l}\left( x_l^{-1} \right) 
 \right].
\end{align}

\section{Summary on the relevant modular forms}
\label{sect:summary_modular_forms}

In this appendix we collect useful formulae for all modular forms of $\Gamma_1(12)$
appearing in the Feynman parameter calculation of the sunrise integral and the kite integral.
For all modular forms we give several equivalent representations: 
A representation in the form of an eta quotient (if it exists), 
a representation in terms of $\mathrm{ELi}$-functions 
and a representation as a linear combination of generalised Eisenstein series. 
We use the notation $\tau_2=\tau/2$ and $q_2=\exp(2 \pi i \tau_2)=\exp(\pi i \tau).$
Let us start with $\psi_1/\pi \in \mathcal{M}_1(12,\chi_1)$.
\begin{align}
 \frac{\psi_1}{\pi} 
 & =
 \frac{2 \mu^2}{\sqrt{3} m^2}
 \frac{\eta\left(\tau_2\right)^3 \eta\left(4 \tau_2\right)^3 \eta\left(6\tau_2\right)}{ \eta\left(2\tau_2\right)^{3} \eta\left(3\tau_2\right) \eta\left(12 \tau_2 \right)}
 \nonumber \\
 & =
 \frac{2 \mu^2}{\sqrt{3} m^2}
 \left[ 1 
  + \frac{\sqrt{3}}{2} \overline{\mathrm{E}}_{0;0}\left(r_3;-1;-q_2\right)
  + \frac{3 \sqrt{3}}{2} \overline{\mathrm{E}}_{0;0}\left(r_3;1;-q_2\right)
 \right] 
 \nonumber \\
 & =
 - \frac{6 \mu^2}{\sqrt{3} m^2}
 \left[ 
 E_1\left(\tau_2;\bar{\chi}_0,\bar{\chi}_1\right)
 - E_1\left(2\tau_2;\bar{\chi}_0,\bar{\chi}_1\right)
 - 2 E_1\left(4\tau_2;\bar{\chi}_0,\bar{\chi}_1\right)
 \right].
\end{align}
In the sunrise integral, three modular forms $f_1 \in \mathcal{M}_1(12,\chi_1)$,
$f_2 \in \mathcal{M}_2(12,\chi_0)$ and
$f_3 \in \mathcal{M}_3(12,\chi_1)$ occur.
We have
\begin{align}
 f_1 
 & =
 \frac{\left(t+3m^2\right)}{2 \sqrt{6} \mu^2} \; \frac{\psi_1}{\pi}
 \nonumber \\
 & =
 \frac{\sqrt{2}}{2} + \sqrt{6} \overline{\mathrm{E}}_{0;0}\left(r_3;1;-q_2\right)
 \nonumber \\
 & =
 -3 \sqrt{2} \left[ E_1(\tau_2,\bar{\chi}_0,\bar{\chi}_1) - 2 E_1(4 \tau_2,\bar{\chi}_0,\bar{\chi}_1) \right],
 \nonumber \\
 f_2 
 & =
 \frac{1}{i \pi} \frac{\psi_1^2}{W} \frac{\left(3t^2-10m^2 t - 9 m^4 \right)}{2 t \left(t-m^2\right) \left(t-9m^2\right)}
 \nonumber \\
 & =
 - \frac{1}{2}
 + 6 \overline{\mathrm{E}}_{0;-1}\left(-1;1;-q_2\right)
 + \overline{\mathrm{E}}_{0;-1}\left(r_3;-1;-q_2\right)
 - 3 \overline{\mathrm{E}}_{0;-1}\left(r_3;1;-q_2\right)
 \nonumber \\
 & =
 14 B_{2,2}(\tau_2) - 4 B_{2,3}(\tau_2) - 8 B_{2,4}(\tau_2) + 10 B_{2,6}(\tau_2) - 4 B_{2,12}(\tau_2),
 \nonumber \\
 f_3 
 & =
 \frac{\mu^2 \psi_1^3}{\pi W^2}
 \;
 \frac{6}{t \left(t-m^2\right)\left(t-9m^2\right)}
 \nonumber \\
 & =
 3 \sqrt{3} 
 \frac{ \eta\left(2\tau_2\right)^{11} \eta\left(6\tau_2\right)^{7} }
      { \eta\left(\tau_2\right)^{5} \eta\left(4\tau_2\right)^{5} \eta\left(3\tau_2\right) \eta\left(12\tau_2\right) }
 \nonumber \\
 & =
 3 \overline{\mathrm{E}}_{0;-2}\left(r_3;-1;-q_2\right)
 \nonumber \\
 & =
 3 \sqrt{3} \left[
 E_3\left(\tau_2;\bar{\chi}_1,\bar{\chi}_0\right)
 + 2 E_3\left(2\tau_2;\bar{\chi}_1,\bar{\chi}_0\right)
 - 8 E_3\left(4\tau_2;\bar{\chi}_1,\bar{\chi}_0\right)
 \right].
\end{align}
In the kite integral we encounter in addition three modular forms $g_{2,0}, g_{2,1}, g_{2,9} \in \mathcal{M}_2(12,\chi_0)$ 
of weight $2$
and two modular forms $g_{3,0}, g_{3,1} \in \mathcal{M}_3(12,\chi_1)$ of weight $3$.
The weight $2$ modular forms are
\begin{align}
 g_{2,0}
 & =
 \frac{1}{i \pi} \frac{\psi_1^2}{W} \frac{1}{t}
 \nonumber \\
 & = 
 \frac{\eta\left(2\tau_2\right)^{10}\eta\left(6\tau_2\right)^{10}}{\eta\left(\tau_2\right)^{4}\eta\left(3\tau_2\right)^{4}\eta\left(4\tau_2\right)^{4}\eta\left(12\tau_2\right)^{4}}
 \nonumber \\
 & =
 1 
 - 4 \overline{\mathrm{E}}_{0;-1}\left(r_3;-1;-q_2\right)
 \nonumber \\
 & =
 4 \left[ B_{2,2}(\tau_2) + B_{2,3}(\tau_2) - B_{2,4}(\tau_2) - B_{2,6}(\tau_2) + B_{2,12}(\tau_2) \right], 
 \nonumber \\
 g_{2,1}
 & =
 \frac{1}{i \pi} \frac{\psi_1^2}{W} \frac{1}{t-m^2}
 \nonumber \\
 & =
 9
 \frac{\eta\left(\tau_2\right)\eta\left(4\tau_2\right)\eta\left(6\tau_2\right)^{12}}{\eta\left(2\tau_2\right)^{4}\eta\left(3\tau_2\right)^{3}\eta\left(12\tau_2\right)^{3}}
 \nonumber \\
 & =
 - \frac{3}{2} \overline{\mathrm{E}}_{0;-1}\left(r_3;-1;-q_2\right)
 + \frac{3}{2} \overline{\mathrm{E}}_{0;-1}\left(r_3;1;-q_2\right)
 + 3 \overline{\mathrm{E}}_{0;-1}\left(-1;1;-q_2\right)
 \nonumber \\
 & =
 3 \left[ 6 B_{2,2}(\tau_2) + B_{2,3}(\tau_2) - 3 B_{2,4}(\tau_2) - 2 B_{2,6}(\tau_2) + B_{2,12}(\tau_2) \right], 
 \nonumber \\
 g_{2,9}
 & =
 \frac{1}{i \pi} \frac{\psi_1^2}{W} \frac{1}{t-9m^2}
 \nonumber \\
 & =
 \frac{\eta\left(2\tau_2\right)^{16}\eta\left(3\tau_2\right)^{5}\eta\left(12\tau_2\right)^{5}}{\eta\left(\tau_2\right)^{7}\eta\left(4\tau_2\right)^{7}\eta\left(6\tau_2\right)^{8}}
 \nonumber \\
 & =
  \frac{1}{2} \overline{\mathrm{E}}_{0;-1}\left(r_3;-1;-q_2\right)
 - \frac{9}{2} \overline{\mathrm{E}}_{0;-1}\left(r_3;1;-q_2\right)
 + 3 \overline{\mathrm{E}}_{0;-1}\left(-1;1;-q_2\right)
 \nonumber \\
 & =
 - 2 B_{2,2}(\tau_2) - 5 B_{2,3}(\tau_2) - B_{2,4}(\tau_2) + 14 B_{2,6}(\tau_2) - 5 B_{2,12}(\tau_2).
\end{align}
For the modular forms of weight $3$ we have the expressions
\begin{align}
 g_{3,0}
 & =
 \frac{1}{i \pi \mu^2} \frac{\psi_1^2}{W} \frac{\psi_1}{\pi}
 \nonumber \\
 & = 
 -6 \sqrt{3} 
 \frac{ \eta\left(2\tau_2\right)^{11} \eta\left(6\tau_2\right)^{7} }
      { \eta\left(\tau_2\right)^{5} \eta\left(4\tau_2\right)^{5} \eta\left(3\tau_2\right) \eta\left(12\tau_2\right) }
 \nonumber \\
 & =
 -6 \overline{\mathrm{E}}_{0;-2}\left(r_3;-1;-q_2\right)
 \nonumber \\
 & =
 -6 \sqrt{3} \left[
 E_3\left(\tau_2;\bar{\chi}_1,\bar{\chi}_0\right)
 + 2 E_3\left(2\tau_2;\bar{\chi}_1,\bar{\chi}_0\right)
 - 8 E_3\left(4\tau_2;\bar{\chi}_1,\bar{\chi}_0\right)
 \right],
 \nonumber \\
 g_{3,1}
 & =
 \frac{1}{i \pi \mu^2} \frac{\psi_1^2}{W} \frac{\psi_1}{\pi} \frac{t}{t-m^2}
 \nonumber \\
 & =
 - 54 \sqrt{3} \frac{\eta\left(6\tau_2\right)^{9}}{\eta\left(2\tau_2\right)^{3}}
 \nonumber \\
 & =
  - \frac{27}{4} \overline{\mathrm{E}}_{0;-2}\left(r_3;-1;-q_2\right)
  - \frac{27}{4} \overline{\mathrm{E}}_{0;-2}\left(r_3;1;-q_2\right)
 \nonumber \\
 & =
 -54 \sqrt{3} E_3\left(2\tau_2;\bar{\chi}_1,\bar{\chi}_0\right).
\end{align}
Note that we have $g_{3,0}=-2f_3$ and
\begin{align}
 f_2 & = 
 - \frac{1}{2} g_{2,0} + g_{2,1} + g_{2,9}.
\end{align}
Furthermore we note that the modular form $g_{3,1}$ is actually already a modular form at level $N=6$.
The values at $\tau_2=i \infty$ for the various modular forms are given by
\begin{align}
 &
 \frac{\psi_1(i \infty)}{\pi} 
 =
 \frac{2 \mu^2}{\sqrt{3} m^2},
 &&&&
 \nonumber \\
 & f_1(i \infty) 
 =
 \frac{1}{2} \sqrt{2},
 & &
 f_2(i \infty) 
 =
 - \frac{1}{2},
 & &
 f_3(i \infty) 
 =
 0,
 \nonumber \\
 & g_{2,0}(i \infty)
 =
 1,
 & &
 g_{2,1}(i \infty)
 =
 0,
 & &
 g_{2,9}(i \infty)
 =
 0,
\nonumber \\
 & g_{3,0}(i \infty)
 =
 0,
 & &
 g_{3,1}(i \infty)
 =
 0.
 & &
\end{align}

\end{appendix}

\bibliography{/home/stefanw/notes/biblio}
\bibliographystyle{/home/stefanw/latex-style/h-physrev5}

\end{document}